\documentclass[final]{siamltex}
\usepackage{epsf}
\usepackage{amssymb}
\usepackage{amsbsy}
\usepackage{verbatim}

\usepackage[sort&compress]{natbib}
\bibpunct{[}{]}{;}{a}{,}{,}
\begin{document}

\title{Stochastic Eulerian-Lagrangian Methods for
Fluid-Structure Interactions with Thermal 
Fluctuations and Shear Boundary Conditions 
}

\author{Paul J. Atzberger
\thanks{University of California, 
Department of Mathematics , Santa Barbara, CA 93106; 
e-mail: atzberg@math.ucsb.edu; phone: 805-893-3239;
Work supported by NSF DMS-0635535.}
}

\maketitle

\begin{abstract}
A computational approach is introduced 
for the study of the rheological properties of 
complex fluids and soft materials.  The approach 
allows for a consistent treatment of 
microstructure elastic mechanics, hydrodynamic coupling, 
thermal fluctuations, and externally driven shear 
flows.  A mixed description in terms of Eulerian and 
Lagrangian reference frames is used for the physical system. 
Microstructure configurations are represented
in a Lagrangian reference frame.  Conserved quantities,
such as momentum of the fluid and microstructures, are 
represented in an Eulerian reference frame.  The 
mathematical formalism couples these different descriptions 
using general operators subject to consistency conditions.  
Thermal fluctuations are taken into account in the formalism
by stochastic driving fields introduced in accordance with 
the principles of statistical mechanics.  To study the 
rheological responses of materials subject to shear, 
generalized periodic boundary conditions are developed
where periodic images are shifted relative to the unit
cell to induce shear.  Stochastic numerical methods are 
developed for the formalism.  As a demonstration of the 
methods, results are presented for the shear responses 
of a polymeric fluid, lipid vesicle fluid, and a gel-like 
material.
\end{abstract}


\pagestyle{myheadings}
\thispagestyle{plain}
\markboth{P.J. ATZBERGER}
{SELM FOR SOFT MATERIALS AND COMPLEX FLUIDS}

%\section*{PREPRINT NOTE:} 
%This preprint is under review and is being 
%actively checked and revised.  As such, the 
%materials presented here may contain errors
%and is subject to revision.  Please feel 
%free to send any comments, 
%errors, or suggestions to 
%atzberg@math.ucsb.edu. \\
%\\
%\textbf{Last updated:} \today; \ampmtime.\\

\section{Introduction} 
Soft materials and complex fluids 
are comprised of microstructures which 
have mechanics and interactions 
characterized by energy scales comparable 
to thermal energy.  This feature results in
interesting bulk material properties and phenomena
which often depend sensitively on temperature and applied 
stresses~\citep{Bird1987Vol1,Barnes1997,Doi1986}.  Example materials 
include colloidal suspensions, foams, polymeric fluids, surfactant 
solutions, lipid vesicles, and gels 
~\citep{Lubensky1997, Hamley2003, bird1987, Gompper2006, Doi1986, Coussot2007, Raub2007}.  
Microstructures of such materials include flexible filaments, bubbles,  
colloidal particles, lipid chains, and polymers.  These
microstructures are typically surrounded by a solvating 
fluid which further mediates interactions through 
solvation shells~\citep{Meyer2006, Leckband2001} and hydrodynamic 
coupling~\citep{Bird1987Vol1, Brady1988, Doi1986}.
In addition, given the energy scales of the microstructure 
mechanics and interactions, thermal fluctuations often play an 
important role both in microstructure organization 
and kinetics~\citep{bird1987, Brady1988, Doi1986}.  
A fundamental challenge in the study of soft materials is 
to relate bulk material properties to microstructure mechanics, 
interactions, and kinetics.

For the study of soft materials we introduce a modeling and simulation 
approach which consistently accounts for microstructure elastic 
mechanics, hydrodynamic coupling, and thermal fluctuations.  
The modeling approach is based on a mixed Eulerian and Lagrangian 
description.  The microstructure configurations are modeled 
in a Lagrangian reference frame, while an Eulerian reference 
frame is used to account for conserved quantities, such as momentum, 
of the system.  When coupling these disparate descriptions an 
important issue is to formulate methods which do not introduce 
artifacts into the conservation laws, such as artificial 
dissipation of energy or loss of momentum.  These properties are 
especially important when introducing stochastic 
driving fields to account for thermal fluctuations.
We discuss a general approach for developing such coupling schemes,
focusing primarily on one such realization referred to as the 
Stochastic Immersed Boundary Method~\citep{Atzberger2007a, Peskin2002}.

To facilitate studies of the rheological properties of soft materials
we introduce methods to account for externally driven 
shear flows.  To account for shearing of the material, we  
generalize the usual periodic boundary
conditions so that periodic images are shifted relative to the 
unit cell to induce shear.  Our approach is based on boundary 
conditions introduced for Molecular Dynamics simulations which are 
referred to as Lees-Edwards boundary conditions~\citep{LeesEdwards1972}.  
These conditions present a number of challenges in the context 
of numerically solving the hydrodynamic equations.  We 
develop numerical methods which utilize jump conditions in 
the velocity field at domain boundaries and utilize a 
change of variable to facilitate handling of the shifted 
boundaries.  Further issues arise when accounting for the 
thermal fluctuations.  For the introduced discretizations 
we develop stochastic driving fields which 
yield stochastic numerical methods which are consistent with 
the principles of statistical mechanics.

We consider primarily two physical regimes.  In the first,
the relaxation dynamics of the fluid modes is explicitly
resolved.  In the second, the fluid modes are treated as
as having relaxed to a quasi-steady-state distribution.  
In the first regime we develop efficient stochastic numerical
methods for the generation of the corresponding fluctuating 
fields of the fluid.  In the second, we develop 
efficient stochastic numerical methods which account for
the correlated stochastic driving fields which account
for the effective thermal fluctuations which drive 
the microstructure dynamics.

As a demonstration of the proposed stochastic numerical
methods, simulations are performed for specific systems.
These include studying the shear responses of 
(i) a polymeric fluid, (ii) a vesicle fluid, and 
(iii) a gel-like material.  To relate microstructure 
interactions and kinetics to bulk material properties 
we develop estimators for an associated macroscopic 
stress tensor.  The estimators take into account the n-body 
interactions in the microstructure mechanics and the 
generalized boundary conditions.  For the polymeric fluid,    
this notion of stress is used to investigate 
the dependence of the shear viscosity and 
normal stresses on the rate of shear.  The vesicle fluid 
is subject to oscillating shear and simulations 
are preformed to characterize the frequency response
in terms of the elastic storage modulus and 
viscous loss modulus over a wide range of frequencies.
As a further demonstration of the methods, the
time dependent shear viscosity of a gel-like material 
is studied through simulations.

The ability to simulate explicitly the microstructure mechanics,
hydrodynamic coupling, and thermal fluctuations provides an 
important link between bulk material properties
and phenomena at the level of the microstructures.  The presented 
framework and related stochastic numerical methods are 
expected to be applicable in the modeling and simulation of 
a wide variety of soft materials and complex fluids.  
The general approaches introduced for coupling the Eulerian and 
Lagrangian descriptions and for the incorporation of thermal 
fluctuations are expected to allow for the development
of many different types of Stochastic Eulerian Lagrangian 
Methods.

\section{Stochastic Eulerian-Lagrangian Modeling Approach}
\begin{figure}[t*]
\centering
\epsfxsize = 5in
\epsffile[14 14 375 178]{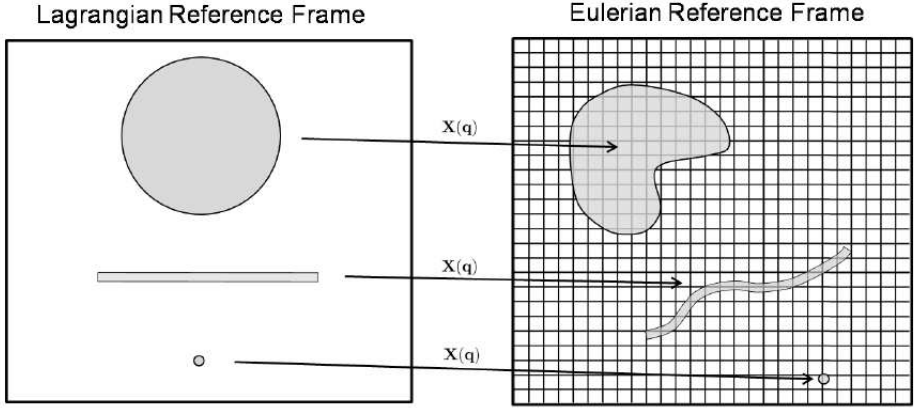}
\caption[Eulerian and Lagrangian Reference Frames]
{A description of the physical system is used which combines
Eulerian and Lagrangian reference frames.
The configuration of the microstructures are described 
using a Lagrangian reference frame, shown on the left.
Microstructures represented in the Lagrangian frame may 
be solid bodies, membrane structures, polymeric structures, 
or point particles.  The conserved quantities, such as the local 
momentum, mass, or energy, are described in an Eulerian reference 
frame, shown on the right.  The mapping $\mathbf{X}(\mathbf{q})$ relates the 
Lagrangian reference frame  to the Eulerian reference frame.
}
\label{figure_EL_schematic}
\end{figure}

We use a mixed Eulerian-Lagrangian description.
The conserved quantities of the entire system including 
both the fluid and microstructures will be accounted for 
in an Eulerian reference frame.  The microstructure 
configurations will be accounted for in a Lagrangian 
reference frame, see Figure~\ref{figure_EL_schematic}.  
To introduce the basic approach and simplify the presentation 
we consider here only a rather special case.  We shall assume 
the solvent fluid is an incompressible 
Newtonian fluid of constant density and the microstructures 
are density matched with the fluid.  In this case, the primary 
conserved quantity of interest is the local momentum of the material.
The basic framework and principles that will be presented are 
more generally applicable allowing for additional conserved quantities 
to be taken into account, such as the local mass density and 
energy~\citep{Donev2009}.  A more abstract and general presentation 
of the formalism will be the focus of another paper.

The basic Eulerian-Lagrangian formalism describing the state of
the fluid and microstructures is given by the following equations
\begin{eqnarray}
\label{equ_EL_E}
\frac{D \mathbf{p}(\mathbf{x},t)}{D t} & = & 
\nabla \cdot \boldsymbol{\sigma}(\mathbf{x},t) 
+ \Lambda(\mathbf{x},t) 
+ \lambda(\mathbf{x},t)
+ \mathbf{g}(\mathbf{x},t) 
\\
\label{equ_EL_L}
\frac{\partial \mathbf{X}(\mathbf{q},t)}{\partial t} & = & \Gamma(\mathbf{q},t) 
+ \gamma(\mathbf{q},t)
+ \mathbf{Z}(\mathbf{q},t).
\end{eqnarray}
The $\mathbf{p}$ accounts for the momentum of the material occupying location $\mathbf{x}$ and 
${D \mathbf{p}}/{D t}$ denotes the material derivative.  The
$\mathbf{X}(\mathbf{q},t)$ denotes the configuration of the material at time $t$ and
parameterized by $\mathbf{q}$.  The local material stress is denoted by 
$\boldsymbol{\sigma} = \boldsymbol{\sigma}[\mathbf{p},\mathbf{X}]$.
We use the convention that $\boldsymbol{\sigma}$ accounts only for the dissipative 
stress contributions in the system.  
The operators $\Lambda$, $\Gamma$ couple 
the Eulerian and Lagrangian descriptions of the state of the material.  
The operator 
$\Lambda = \Lambda[\mathbf{X}]$ accounts 
for momentum gained or lost locally in the system
as the material deforms from non-dissipative stresses and body forces.  
The operator 
$\Gamma = \Gamma[\mathbf{p}, \mathbf{X}]$ determines the rate of deformation of the material
from the momentum of the system.
The $\lambda = \lambda[\mathbf{X},\mathbf{p}]$ and $\gamma = \gamma[\mathbf{X},\mathbf{p}]$ 
are Lagrange multipliers associated with time-independent kinematic constraints imposed on the system, 
such as rigidity of a body or incompressibility.  Thermal fluctuations are taken into 
account through the stochastic fields $\mathbf{g}$ and $\mathbf{Z}$.  

We consider systems where the total energy is given by
\begin{eqnarray}
\label{equ_SELM_energy}
E[\mathbf{p}, \mathbf{X}] = \int \frac{\rho_0}{2}|\mathbf{u}(\mathbf{x})|^2 d\mathbf{x} + \Phi[\mathbf{X}],
\end{eqnarray}
where $\mathbf{u}(\mathbf{x}) = \rho_0^{-1}\mathbf{p}(\mathbf{x})$ is the velocity 
of the material at location $\mathbf{x}$, $\rho_0$ is the constant mass density of 
the material, and $\Phi$ is the potential energy for a given configuration.  The 
force associated with this energy is denoted by 
$\mathbf{F} = -\delta \Phi / \delta \mathbf{X}$.

For the operators which couple the Eulerian 
and Lagrangian descriptions to be physically consistent,
the following should hold: (i) the coupling 
operators should not introduce any loss or gain of energy, (ii) 
momentum should only change through forces acting within 
the system.  More precisely, 
these conditions require
\begin{eqnarray}
\label{equ_coupling_cond_energy}
\int \mathbf{F}(\mathbf{q})\cdot \Gamma(\mathbf{q}) d\mathbf{q}
& = &
\int
\left[\rho_0^{-1}\mathbf{p}(\mathbf{x})\right]
\cdot
\Lambda[\mathbf{F}](\mathbf{x})
d\mathbf{x}
\\
\label{equ_coupling_cond_momentum}
\int_{\Omega} \Lambda[\mathbf{F}](\mathbf{x}) d\mathbf{x} & = & 
\int \mathbf{F}(\mathbf{q}) d\mathbf{q}.
\end{eqnarray}
The conditions are required to hold for any realization of $\mathbf{X}$, 
$\mathbf{p}$, and $\mathbf{F}$.  The condition~\ref{equ_coupling_cond_energy} 
ensures the coupling operators conserve energy.  The 
condition~\ref{equ_coupling_cond_momentum} ensures that in 
the absence of constraints the total momentum change of the system 
is equal to the total force acting on the system.

In the notation, we find it convenient
to write the operator $\Lambda$ as explicitly depending on both 
$\mathbf{X}$ and $\mathbf{F}$, which for conservative forces is 
technically redundant.  To simplify 
the discussion, it has been 
assumed that the stress contributions denoted by $\boldsymbol{\sigma}$ 
are entirely dissipative and that there is no net in-flux of momentum 
from boundary stresses 
$\int_{\partial \Omega} \boldsymbol{\sigma}(\mathbf{x})\cdot \mathbf{n} d\mathbf{x} = 0$.

With these conditions satisfied by the coupling operators, 
we discuss how to account for thermal fluctuations using 
the stochastic fields $\mathbf{g}$ and $\mathbf{Z}$.
It is convenient when accounting for thermal 
fluctuations to introduce coupling operators so that 
all configurations $\mathbf{X}$ are equally probable 
at statistical steady-state, when the 
$\Phi[\mathbf{X}] \equiv 0$.  It can be shown that this corresponds
to dynamics determined by the constraints and coupling operators 
which introduces an incompressible flow on phase space.  The 
requirement of an incompressibile flow on phase space can 
be expressed as
\begin{eqnarray}
\label{equ_coupling_cond_uniform}
\int 
\frac{\delta \Lambda}{\delta\mathbf{p}}(\mathbf{x},\mathbf{x}) 
d\mathbf{x}
+
\int 
\frac{\delta \lambda}{\delta\mathbf{p}}(\mathbf{x},\mathbf{x}) 
d\mathbf{x}
+
\int
\frac{\delta \Gamma}{\delta\mathbf{X}}(\mathbf{q},\mathbf{q})
d\mathbf{q}
+
\int 
\frac{\delta \gamma}{\delta\mathbf{X}}(\mathbf{q},\mathbf{q})
d\mathbf{q} & = & 0.
\end{eqnarray}
This condition can be relaxed to allow for more general 
choices of coordinates, coupling operators, and constraints.
If this condition is not satisfied a more general treatment 
of the thermal fluctuations is required to take into account
in the invariant distribution the local compression or 
dilation of volume under the phase space flow~\citep{Tuckerman1999}.

To simplify the discussion, we assume that the dissipative 
processes can be accounted for by a negative definite 
self-adjoint linear operator $\mathcal{L}$ in $\mathbf{p}$, 
so that $\nabla\cdot \boldsymbol{\sigma} = \mathcal{L}\mathbf{p}$, and that 
conditions \ref{equ_coupling_cond_energy} - \ref{equ_coupling_cond_uniform}
are satisfied.  With these assumptions, the thermal fluctuations can be accounted 
for using for $\mathbf{g}$ and $\mathbf{Z}$ 
Gaussian stochastic fields which are mean zero and $\delta$-correlated in 
time~\citep{Oksendal2000,Gardiner1985}.
The main issue then becomes to determine an 
appropriate spatial covariance structure for these stochastic fields.  
By requiring that the Boltzmann distribution be invariant under the stochastic 
dynamics of equations \ref{equ_EL_E} - \ref{equ_EL_L}, it is required that
$\mathbf{Z} = 0$, and that 
\begin{eqnarray}
\label{equ_SELM_covOp_G}
G(\mathbf{x},t,\mathbf{y},s) = \langle(\mathbf{g}(\mathbf{x},t))(\mathbf{g}(\mathbf{y},s))^{T}\rangle 
= -2k_B {T} \rho_0 \delta(t - s)\mathcal{L} \delta(\mathbf{x} - \mathbf{y}),
\end{eqnarray}
see Appendix~\ref{appendix_invariance_Boltzmann}.  

Similar formulations as equations \ref{equ_EL_E} - \ref{equ_EL_L},
with $\mathbf{g} = 0$, $\mathbf{Z} = 0$, are the starting point 
for the derivation of a wide variety of computational approaches 
used for systems in which fluids interact with rigid or elastic bodies.
These include Arbitrary Lagrangian-Eulerian Methods (ALE)~\citep{Donea1983, Donea2004},
Fluctuating Immersed Material (FIMAT) Dynamics~\citep{Patankar2004},
Immersed Finite Element Methods (IFEM)~\citep{Wang2009, Zhang2004}, 
and Immersed Boundary Methods (IBM)~\citep{Atzberger2007a, Peskin2002}.
The approaches we introduce allow for the incorporation 
and simulation of thermal fluctuations in such methods, which 
collectively we refer to as Stochastic Eulerian-Lagrangian 
Methods (SELMs).

\section{Semi-Discretization of the Momentum Equations, Microstructures, and 
the Eulerian-Lagrangian Coupling}
We now consider semi-discretizations of the SELM equations.
The momentum equations will be spatially discretized on a uniform 
mesh.  The $\mathbf{p}_{\mathbf{m}}$ will denote the momentum at 
the mesh site indexed by $\mathbf{m} = (m_1,m_2,m_3)$ and the 
composite vector of such values will be denoted by $\mathbf{p}$.
The deformation state which describes the microstructure configurations
will be discretized using a finite number of degrees of freedom
denoted by $\mathbf{X}^{[j]}$ indexed by $j = 0,1,\ldots,M$ and 
the composite vector denoted by $\mathbf{X}$.  As an energy for 
this discretized system we use
\begin{eqnarray}
\label{equ_SELM_energy_discr}
E[\mathbf{p},\mathbf{X}] 
= \sum_{\mathbf{m}} \frac{1}{2} \rho_{0}^{-1} |\mathbf{p}_{\mathbf{m}}|^2 \Delta{x}^d 
+ \Phi(\mathbf{X})
\end{eqnarray}
where $\Delta{x}$ is the mesh spacing and $d$ is the number of dimensions.
The semi-discretization in space of the momentum and configuration equations 
can be expressed as
\begin{eqnarray}
\label{equ_EL_E_discr}
\frac{\tilde{D} \mathbf{p}}{\tilde{D} t} & = & L \mathbf{p} + \Lambda + \lambda + \mathbf{g} \\
\label{equ_EL_L_discr}
\frac{\partial \mathbf{X}^{[j]}(t)}{\partial t} & = & \Gamma^{[j]} + \gamma^{[j]}
\end{eqnarray}
where ${\tilde{D}}/{\tilde{D}t}$ and $L$ denote respectively the 
spatially discretized approximation of the 
material derivative and $\mathcal{L}$.  The
$\mathbf{p}$, $\Lambda$, $\lambda$, $\mathbf{g}$ denote the composite vector of values on
the mesh and $\mathbf{X}^{[j]}$, $\Gamma^{[j]}$, $\gamma^{[j]}$ denote values associated with the 
$j^{th}$ configurational degree of freedom.  We assume the discrete dissipative 
operator is symmetric $L = L^T$ and negative semi-definite.
For the coupling operators of the discretized equations the corresponding consistency 
conditions~\ref{equ_coupling_cond_energy} - \ref{equ_coupling_cond_momentum} 
are
\begin{eqnarray}
\label{equ_coupling_cond_energy_discr}
\Gamma[\mathbf{p}]^T \mathbf{F} 
& = & 
\rho_0^{-1}
\mathbf{p}^T
\Lambda[\mathbf{F}] \Delta{x}^d \\
\label{equ_coupling_cond_momentum_discr}
\sum_{\mathbf{m}} \Lambda[\mathbf{F}]_{\mathbf{m}} \Delta{x}^d 
& = & \sum_{j} \mathbf{F}^{[j]}.
\end{eqnarray}
The superscript T denotes the matrix transpose. 
The first condition ensures for the discretized system that 
the coupling preserves the energy and the second that 
changes in momentum only occur from forces acting
within the system.  The phase space incompressibility condition
corresponding to equation \ref{equ_coupling_cond_uniform} 
in the discretized setting becomes
\begin{eqnarray}
\label{equ_coupling_cond_uniform_discr}
\nabla_{\mathbf{p}} \cdot (\Lambda + \lambda)
+ \nabla_{\mathbf{X}} \cdot (\Gamma + \gamma) & = & 0.
\end{eqnarray}
This condition ensures the uniform distribution 
for the configurations $\mathbf{X}$ is invariant under the stochastic 
dynamics of equation \ref{equ_EL_E_discr}-\ref{equ_EL_L_discr} when 
the potential energy is constant, i.e. $\Phi \equiv 0$.
When $\Lambda$ and $\Gamma$ are linear operators the energy 
condition~\ref{equ_coupling_cond_energy} amounts to the 
coupling operators being adjoints (up to a scalar), 
\begin{eqnarray}
\Gamma = \Lambda^T\rho_0\Delta{x}^d.  
\end{eqnarray}

Provided conditions~\ref{equ_coupling_cond_energy_discr}~-~\ref{equ_coupling_cond_uniform_discr} 
are satisfied, the thermal fluctuations can be taken into account 
using a Gaussian stochastic field on the lattice, without any direct
stochastic forcing required in the microstructure equations~\ref{equ_EL_L}.
If these conditions are violated by the discretization the numerical 
approximation may introduce artificial loss or gain of energy or momentum in 
the system.  In order to be consistent with the fluctuation-dissipation principle
of statistical mechanics such discretizations would require additional 
sources of stochastic forcing to obtain the appropriate Boltzmann 
ensemble.  

An appropriate covariance structure for the stochastic
driving field for discretizations satisfying 
conditions~\ref{equ_coupling_cond_energy_discr}~-~\ref{equ_coupling_cond_uniform_discr}
can be determined by 
requiring invariance of the Boltzmann distribution 
under the stochastic dynamics of 
equation~\ref{equ_EL_E_discr}~-~\ref{equ_EL_L_discr}.
This yields for the semi-discrete system, see Appendix~\ref{appendix_invariance_Boltzmann},
\begin{eqnarray}
\label{equ_SELM_covOp_G_discr}
G = \langle\mathbf{g}\mathbf{g}^T\rangle 
= -2LC.
\end{eqnarray}
The covariance of the equilibrium fluctuations is given by the entries
\begin{eqnarray}
C = 
\frac{\rho_0{k_B {T}}}{\Delta{x}^d} I,
\end{eqnarray}
where $I$ is the identity matrix, see Appendix~\ref{appendix_invariance_Boltzmann}.

One such realization of this SELM approach is the 
Stochastic Immersed Boundary Method developed 
in~\citep{Atzberger2007a}.  In this case the coupling 
operators are given by 
\begin{eqnarray}
\label{equ_op_SIB_Lambda}
\Lambda_{IB} \mathbf{F} & = & 
\sum_{j = 1}^{M}\mathbf{F}^{[j]}(\mathbf{X}(t))\delta_{a}(\mathbf{x}_{\mathbf{m}} - \mathbf{X}^{[j]}(t))\\
\label{equ_op_SIB_Gamma}
\left[\Gamma_{IB} \mathbf{u}\right]^{[j]}
& = & 
\sum_{\mathbf{m}} \delta_a(\mathbf{x}_{\mathbf{m}} - \mathbf{X}^{[j]}(t))
\mathbf{u}_{\mathbf{m}}(t) \Delta{x}^d.
\end{eqnarray}
where $\mathbf{u} = \rho_0^{-1}\mathbf{p}$, and $\delta_a$ is a special kernel
function approximating the Dirac $\delta$-function, see 
Appendix~\ref{appendix_delta_func}.  The 
dynamics are subject to the constraint that the fluid is incompressible 
$\nabla\cdot\mathbf{u} = 0$.  For the semi-discretized 
SIB method of~\citep{Atzberger2007a} the 
conditions~\ref{equ_coupling_cond_energy_discr} and \ref{equ_coupling_cond_momentum_discr} 
can be readily verified to hold exactly.  However, the 
condition~\ref{equ_coupling_cond_uniform_discr} only approximately holds
and is exact only in the continuum limit.  As
$\Delta{x} \rightarrow 0$ we have
\begin{eqnarray}
\nabla_{\mathbf{X}^{[j]}} \cdot \Gamma
& = & \sum_{\mathbf{m}} -\nabla\delta_a(\mathbf{x}_{\mathbf{m}} - \mathbf{X}^{[j]})
\mathbf{u}_{\mathbf{m}} \Delta{x}^d  
\rightarrow  \int -\nabla\delta_a(\mathbf{y} - \mathbf{X}^{[j]}) \mathbf{u}({\mathbf{y}}) d\mathbf{y}\\
\nonumber
& = & \int \delta_a(\mathbf{y} - \mathbf{X}^{[j]})
\nabla\cdot\mathbf{u}({\mathbf{y}}) d\mathbf{y} = 0.
\end{eqnarray}
In the last line we used that the fluid is incompressible $\nabla\cdot \mathbf{u} = 0$.
The divergence of the terms $\Lambda$, $\lambda$, $\gamma$, are 
zero in this case.  In practice, the method still yields reasonable 
results since the exhibited fluctuations deviate from Boltzmann 
statistics only up to the discretization error~\citep{Atzberger2007a, Atzberger2008}.

\section{Soft Materials and Complex Fluids Subject to Shear}
\label{sec_soft_materials_subject_shear}
We now discuss how the SELM approach can be used for 
the study of rheological properties of 
soft materials and complex fluids.  We then discuss
specific stochastic numerical methods for performing
simulations in practice.  For the type of materials 
we consider, it will be assumed that the 
solvent hydrodynamics is described well by the 
constitutive laws of Newtonian fluids in the 
physical regime where the Reynolds 
number is small.  We also assume that the 
explicitly represented microstructures occupy 
only a relatively small 
volume fraction and are effectively 
density matched with the solvent fluid.  In this 
regime, the momentum of the system will be accounted 
for using the time-dependent Stokes equations
\begin{eqnarray}
\label{equ_time_dep_stokes}
\rho_0{\frac{\partial  \mathbf{u}}{\partial t}} & = & 
\mu{\Delta}{\mathbf{u}} - \nabla{p} 
+ \Lambda + \mathbf{g} \\
\nabla\cdot{\mathbf{u}} & = & 0 
\end{eqnarray}
where $\mathbf{u}(\mathbf{x},t) = \rho_0^{-1}\mathbf{p}(\mathbf{x},t)$ is 
the local velocity of the fluid body at $\mathbf{x}$ in 
the Eulerian reference frame, $\rho_0$ is the fluid density, 
$\mu$ is the dynamic viscosity, and $p$ is
the pressure. This corresponds to the dissipative stress 
$\boldsymbol{\sigma} = \mu\left(\nabla\mathbf{u} + \nabla\mathbf{u}^T\right)$
and $\lambda = -\nabla{p}$ in equation~\ref{equ_EL_E}.  
While the Reynolds number is small, the partial time 
derivative is retained in the Stokes flow 
in equation \ref{equ_time_dep_stokes}
since the thermal fluctuations 
introduce small characteristic time scales into the dynamics.

\begin{figure}[t*]
\centering
\epsfxsize = 5in
\epsffile[14 14 375 157]{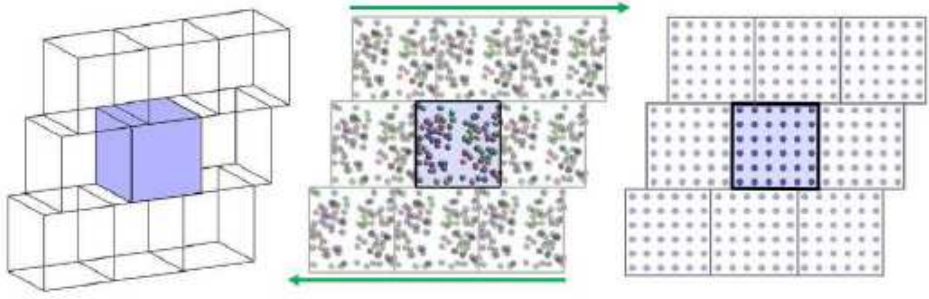}
\caption[Discretization Mesh with Lees-Edwards Boundary Conditions]
{Discretization Mesh with Lees-Edwards Boundary Conditions Initial Formulation.
The boundary conditions induce shear by shifting the periodic images of
the unit cell, shown on the left and middle.  For uniform discretizations 
of the unit cell this presents challenges since the mesh becomes misaligned 
at boundaries between the unit cell and the periodic images, shown on the right.}
\label{figure_meshLeesEdwardsI}
\end{figure}

To introduce shear we generalize the usual periodic 
boundary conditions.  Our basic approach is motivated by
the molecular dynamics methods introduced by 
Lees-Edwards~\citep{LeesEdwards1972,Evans1979,EvansMorriss1984}.  
In this work, molecules in the base unit cell have 
modified interactions with molecules in periodic 
images.  To simulate a bulk material
undergoing a shear deformation at a given rate, the 
periodic images are treated as shifting in time relative
to the unit cell, see Figure~\ref{figure_meshLeesEdwardsI}.  This has the effect
of modifying both the location of periodic images of 
molecules and their assigned velocities.  This 
has some advantages over other approaches,
where an affine-like deformation is imposed on 
the entire material body~\citep{EvansMorriss1990,Hoover1980,Hoover2008}.  
In contrast, for the Lees-Edwards approach the shear deformation 
is only imposed at the boundaries allowing within the unit 
cell for the molecular interactions to determine the form 
of the shear response.

Motivated by this molecular dynamics condition we  
develop a corresponding methodology for the SELM approach.  For momentum 
accounted for by the time-dependent Stokes equations 
we introduce the following generalized periodic boundary conditions
\begin{eqnarray}
\label{equ_StokesJumpBndCond}
\mathbf{u}(x,y,L,t) = \mathbf{u}(x - vt,y,0,t) + v\mathbf{e}_{x}.
\end{eqnarray}
For concreteness we  consider the case where a shear is imposed in the
z-direction giving rise to velocities in the x-direction.  The 
$L$ is the side length of the periodic cell in the z-direction,
$v = L\dot{\gamma}$ is the velocity of the top face of the unit cell 
relative to the bottom face,  $\dot{\gamma}$ denotes the rate of 
shear deformation, and $\mathbf{e}_j$ is the standard unit vector in the $j^{th}$
direction.  The interactions between microstructures of the system can be readily handled
in the same manner as in the molecular dynamics simulation.  This is done by 
shifting the location of any microstructure of a periodic image involved in
an interaction.

While conceptually straight-forward, these boundary conditions present 
significant challenges in practice for the numerical discretization of 
the momentum equations.  The conditions introduce both a jump
discontinuity at periodic boundaries and a shift which potentially 
leads to misalignment of discretization nodes at the domain boundaries,
see Figure~\ref{figure_meshLeesEdwardsI}.
For commonly employed approaches such as spectral Fourier methods 
the jump discontinuity results in a degradation of accuracy through 
the resulting Gibbs' phenomena~\citep{Goldberg1976}.  For uniform finite difference methods 
on the unit cell the mesh misalignment requires modified stencils or 
interpolations at the domain boundary.  When incorporating stochastic 
driving fields to account for thermal fluctuations these issues are 
further compounded.

\begin{figure}[t*]
\centering
\epsfxsize = 5in
\epsffile[14 14 375 143]{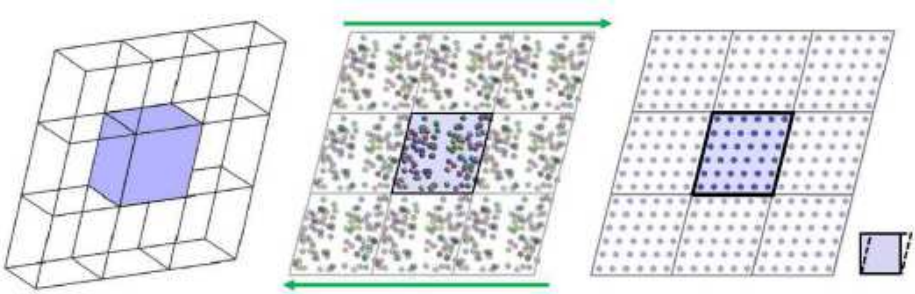}
\caption[Discretization Mesh with Lees-Edwards Boundary Conditions]
{Discretization Mesh with Lees-Edwards Boundary Conditions Using a Moving Coordinate Frame.
By discretizing the momentum equations in a moving coordinate frame a uniform discretization
is obtained in which the mesh of the unit cell aligns with the mesh of the periodic images,
shown on the left, middle, and right.  The definition of the unit cell is changed from a 
cube to a sheared parallelepiped, shown on the far right on the bottom.
}
\label{figure_meshLeesEdwardsII}
\end{figure}

To address these issues, we develop discretization
methods which utilize a moving coordinate frame which 
deforms with the unit cell, see Figure~\ref{figure_meshLeesEdwardsII}.  
Let the velocity 
field in this frame be denoted by 
$\mathbf{w}(\mathbf{q},t) := \mathbf{u}(\phi(\mathbf{q},t),t)$,
where $\mathbf{q} = (q_1,q_2,q_3)$ parameterizes the deformed unit
cell.  Let $\phi(\mathbf{q},t) = (q_1 + q_3\dot{\gamma}t, q_2, q_3)$ 
denote the map from the moving coordinate frame to the fixed 
Eulerian coordinate frame $\mathbf{x} = \phi(\mathbf{q})$.  The time-dependent Stokes equations 
in the deforming coordinate frame become
\begin{eqnarray}
\label{equ_stokesShearFrame_momentum}
{\frac{d\mathbf{w}^{(d)}}{dt}} & = & 
\rho_0^{-1}\mu
\left[
\mathbf{e}_d
-
\delta_{d,3}
\dot{\gamma}t
\mathbf{e}_x
\right]^T
\nabla^2 \mathbf{w}^{(d)}
\left[
\mathbf{e}_d
-
\delta_{d,3}
\dot{\gamma}t
\mathbf{e}_x
\right]
- \nabla{p} 
+ \mathbf{F} + \mathbf{J} \\
\label{equ_stokesShearFrame_continuity}
\nabla
\cdot\mathbf{w}
& - &
\mathbf{e}_z^T
\hspace{0.1cm}
\nabla\mathbf{w}
\hspace{0.1cm}
\mathbf{e}_x 
\dot{\gamma}t
 =  
\mathbf{K}
\end{eqnarray}
where $\mathbf{q} = (q_1,q_2,q_3)$ parameterizes the deformed unit cell,
$\dot{\gamma}$ denotes the rate of the shear deformation, $\mathbf{e}_i$ 
the standard basis vector in the $i$ direction with $i \in \{x,y,z\}$.
In the notation the parenthesized superscript denotes a vector component and
$\delta_{k,\ell}$ denotes the Kronecker $\delta$-function.  
We also use the notational convention 
\begin{eqnarray}
\label{equ_components_Laplace_w}
\left[\nabla^2 \mathbf{w}^{(d)}\right]_{i,j} 
&=& 
\frac{\partial^2 \mathbf{w}^{(d)}}{\partial q_i \partial q_j}\\
\label{equ_components_grad_w}
\left[\nabla \mathbf{w}\right]_{d,j} 
&=& 
\frac{\partial \mathbf{w}^{(d)}}{\partial q_j}.
\end{eqnarray}
In the equations, the terms $\mathbf{J},\mathbf{K}$ are introduced 
to account for the jump introduced by the boundary 
conditions~\ref{equ_StokesJumpBndCond}.  This allows in the 
new coordinate frame for use of the usual periodic boundary 
conditions
\begin{eqnarray}
\mathbf{w}(q_1,q_2,L,t) = \mathbf{w}(q_1,q_2,0,t).
\end{eqnarray}

We now discuss a discretization for equations~\ref{equ_stokesShearFrame_momentum} 
and~\ref{equ_stokesShearFrame_continuity} and the corresponding source terms
$\mathbf{J},\mathbf{K}$.  The following central finite difference approximations
will be used
\begin{eqnarray}
\label{equ_discrFirstDeriv}
\frac{\partial \mathbf{w}^{(d)}}{\partial q_i} 
&\rightarrow & \frac{\mathbf{w}^{(d)}(\mathbf{q} + \mathbf{e}_i)
- \mathbf{w}^{(d)}(\mathbf{q} - \mathbf{e}_i)}{2\Delta{x}} 
\\
\label{equ_discrMixedDeriv}
\frac{\partial^2 \mathbf{w}^{(d)}}{\partial q_i \partial q_j} 
&\rightarrow & 
\frac{
\mathbf{w}^{(d)}(\mathbf{q} + \mathbf{e}_i + \mathbf{e}_j)
-
\mathbf{w}^{(d)}(\mathbf{q} - \mathbf{e}_i + \mathbf{e}_j)
}{4\Delta{x}^2}  \\
\nonumber
& - &
\frac{
\mathbf{w}^{(d)}(\mathbf{q} + \mathbf{e}_i - \mathbf{e}_j)
-
\mathbf{w}^{(d)}(\mathbf{q} - \mathbf{e}_i - \mathbf{e}_j)
}{4\Delta{x}^2}, \mbox{ $i \not= j$} \\
\label{equ_discrSecondDeriv}
\frac{\partial^2 \mathbf{w}^{(d)}}{\partial q_i^2} 
&\rightarrow & 
\frac{
\mathbf{w}^{(d)}(\mathbf{q} + \mathbf{e}_i)
-
2\mathbf{w}^{(d)}(\mathbf{q})
+
\mathbf{w}^{(d)}(\mathbf{q} - \mathbf{e}_i)
}{\Delta{x}^2}. 
\end{eqnarray}
These approximations are substituted into 
equations~\ref{equ_components_Laplace_w}--\ref{equ_components_grad_w}
to approximate the operators in 
equation~\ref{equ_stokesShearFrame_momentum}--\ref{equ_stokesShearFrame_continuity}.

We remark that the moving coordinate frame makes the description of the momentum 
field have some features of a Lagrangian frame of reference.  We none-the-less
retain the Eulerian terminology treating this distinction loosely since the 
deformation corresponds to a somewhat arbitrary coordinate frame 
introduced for numerical convenience and does not directly follow 
from the details of the fluid flow.  Our discretization approach 
shares features with Arbitrary Eulerian Lagrangian (ALE) 
Methods~\citep{Donea1983, Donea2004}.

An important issue when using such  
deforming reference frames is that the 
discretization stencils may become 
excessively distorted~\citep{Donea1983, Donea2004}.  
We avoid this issue by exploiting the periodic
symmetry of the system in the $x$ and
$y$ directions.  Let the displacement
in the $x$-direction of the top of 
the unit cell relative to the bottom
of the unit cell be denoted by 
$s$.  For shear rate $\dot{\gamma}$
and cell size $L$ the displacement
at time $t$ is given by 
$s = L\dot{\gamma}t$.  The periodicity
in the $x$ and $y$ directions
has the consequence that for any
coordinate frame with 
$s > L$ there 
is another coordinate frame with 
$s < L$ which has aligned mesh 
sites, see Figure~\ref{figure_meshLeesEdwardsII}.  
By adopting the convention that
the coordinate frame with 
$s < L$ is always used when 
evaluating stencils the 
distortion is controlled.

To obtain approximations for the source terms 
$\mathbf{J}, \mathbf{K}$ the discretization stencils
are applied at the shear boundaries of the unit cell.
For any stencil weights involving values at 
mesh sites which cross the boundary the modified
image value is used $\mathbf{w}_{\mathbf{m}} \pm \dot{\gamma}L$.
The contributions of the stencil weights multiplied by $\pm \dot{\gamma}L$ 
are collected over all boundary mesh sites to obtain the source 
terms $\mathbf{J}$, $\mathbf{K}$.  This allows for the usual finite difference
stencils to be used on the unit cell with regular periodic boundary conditions.
When including the source terms this gives the equivalent result of
imposing the jump boundary condition~\ref{equ_StokesJumpBndCond}.
This formulation has a number 
of advantages when numerically solving the discretized
equations and when introducing thermal fluctuations.

The Stokes equations~\ref{equ_stokesShearFrame_momentum}--
\ref{equ_stokesShearFrame_continuity}
discretized in this manner on a uniform periodic 
mesh can be expressed as
\begin{eqnarray}
\label{equ_dw_dt_time_op}
{\frac{\partial \mathbf{w}}{\partial t}} & = & 
L(t)\mathbf{w}
+ \mathbf{F} + \mathbf{J} + \mathbf{g}  \\
\label{equ_incomp_time_op}
D(t)\mathbf{w} & = & \mathbf{K}
\end{eqnarray}
where $L(t)$ denotes the finite difference operator
approximating the Laplacian in the moving coordinate
frame, and $D(t)$ the approximation of the Divergence 
operator in the moving frame, see 
equation~\ref{equ_stokesShearFrame_continuity}.  
The discretization $L(t)$ can be shown to be 
symmetric and negative semi-definite for each $t$.

An important property of $L(t), D(t)$ is that 
for each time $t$ the corresponding 
stencils are translation invariant with
respect to lattice shifts of the mesh.
This has the important consequence that the 
matrix representations are circulant
and therefore diagonalizable by Fast Fourier
Transforms~\citep{Strang1988}. %(cite) (Tukey, etc...)
As a result, the incompressibility constraint 
can be handled using FFTs to obtain 
an exact projection method~\citep{Chorin1968}.  
This allows for the discretized approximation of the 
Stokes equations to be expressed as
\begin{eqnarray}
\label{equ_dw_dt}
{\frac{\partial \mathbf{w}}{\partial t}} & = & 
\wp(t) \left[L(t)\mathbf{w}
+ \mathbf{F} +  \mathbf{J} \right] + \mathbf{g}
\end{eqnarray}
where $\wp(t)$ is the operator which projects to the 
null space of $D(t)$.  The incompressibility
condition is then satisfied for all time 
provided $D(0) \cdot \mathbf{w}(0) =  \mathbf{K}$.

We discuss stochastic numerical methods 
for two particular physical regimes: (i) the
relaxation of the hydrodynamic modes of the 
system is resolved explicitly, (ii) for the 
current configuration $\mathbf{X}$ the 
hydrodynamic modes are treated as having 
relaxed to statistical steady-state.  We remark
that the case of resolving the hydrodynamic
relaxation of the system is amenable to 
stochastic numerical methods similar to those
introduced in~\citep{Atzberger2007a}.  We 
discuss this case only briefly and focus primarily 
on the newly introduced stochastic numerical methods 
for handling the second case.

\section{Regime I : Resolution of Hydrodynamic Relaxation} 
\label{section_fluct_dissip}
We now discuss in practice how the stochastic fields 
may be generated in the regime where the relaxation
of the hydrodynamic modes is resolved explicitly.  
For this purpose we express equation~\ref{equ_dw_dt}
in differential form 
\begin{eqnarray}
\label{equ_timeDependStokesStoch}
d\mathbf{w} & = & 
\wp(t) \left[L(t)\mathbf{w}
+ \mathbf{F} +  \mathbf{J} \right]dt 
+ Q d\mathbf{B}_t.
\end{eqnarray}
The $Q d\mathbf{B}_t$ denotes the stochastic
driving field accounting for thermal fluctuations
corresponding to $\mathbf{g}$ and $\mathbf{B}_t \in \mathbb{R}^{3N}$ 
denotes the composite vector of a standard Brownian motion
process at each of the mesh sites.  Throughout our discussion
the stochastic differential equations will be given 
the Ito interpretation~\citep{Gardiner1985}.

Using $\langle Q d\mathbf{B}_t d\mathbf{B}_t^T Q^T \rangle
= QQ^Tdt = G dt$, we see that $Q$ denotes a matrix
square-root of the covariance of the 
stochastic driving field $G = QQ^T$.
Given the discretizations introduced in 
Section~\ref{sec_soft_materials_subject_shear},
the dissipative operator $L(t)$ depends on time, 
which requires, see Appendix~\ref{appendix_fluct_dissip},
\begin{eqnarray}
\label{equ_time_G}
G(t) = -2\wp(t)L(t)C.
\end{eqnarray}
This has the consequence that the 
covariance of the stochastic driving 
field is time dependent.

In the discretized system 
the numerical stencils dependent
on time,   However, since the shear
deformation is volume preserving the
discretized summation introduced to model
the kinetic energy of the discrete system
in equation~\ref{equ_SELM_energy_discr} evaluated
in the deformed coordinates is in fact not 
dependent on time.  Since we made this choice, 
we have the important consequence that the 
Boltzmann equilibrium fluctuations of the 
velocity field $\mathbf{w}$ associated
with this energy are stationary
(independent of time).  In other words,
the covariance $C$ of the equilibrium 
fluctuations on the 
discretized lattice for the energy given in
equation~\ref{equ_SELM_energy_discr}
is independent of time, $C(t) = C(0)$.  This
holds even though the 
underlying discretization and corresponding
operators $L(t)$, $\wp(t)$, $G(t)$ 
depend on time. 

To obtain an explicit form for $G(t)$ we need
to compute $C$ taking into account 
the incompressibility constraint 
\ref{equ_incomp_time_op}.  The equilibrium 
covariance under these constraints is given by
\begin{eqnarray}
\label{equ_barC}
C = \frac{2}{3}\frac{k_B{T}}{\rho_0 \Delta{x}^d}I.
\end{eqnarray}
The factor ${2}/{3}$ arises from application 
in Fourier space of the projection operator which
equivalently enforces the incompressibility.
The factor $\rho_0$ appears in the denominator
since the velocity $\mathbf{w}$ is considered,
instead of the momentum 
$\mathbf{p} = \rho_0 \mathbf{w}$.  The notation
for the stochastic driving field $\mathbf{g}$ 
is used loosely when switching between the 
momentum and velocity equations.

The time dependent covariance structure of the 
stochastic driving field $\mathbf{g}$ in 
equation~\ref{equ_timeDependStokesStoch}
is of the form 
$G(t) = -2\wp(t){L}(t)C$.
An important issue 
is whether this will indeed yield a consistent 
treatment of the thermal fluctuations so that 
the resulting stochastic dynamical system has 
the required equilibrium fluctuations.  We 
establish a Fluctuation-Dissipation principle 
for such time dependent systems in 
Appendix~\ref{appendix_fluct_dissip}.

\section{Generating the Stochastic Driving Field I} 
\label{sec_gen_stoch_I}
In order for equation~\ref{equ_time_G} 
and~\ref{equ_barC} to be useful 
in practice, we must have efficient methods 
by which to generate the stochastic driving 
fields with the required covariance structure. 
A significant challenge in practice is to generate efficiently the 
Gaussian stochastic driving field with 
the required covariance structure $G(t)$.  A commonly used approach 
is to generate a variate with uncorrelated standard Gaussian 
components $\boldsymbol{\xi}$ and set 
$\mathbf{g} = Q(t) \boldsymbol{\xi}$ for an appropriately
chosen matrix $Q(t)$.  The resulting variate $\mathbf{g}$
then has covariance $\langle\mathbf{g}\mathbf{g}^T\rangle
=  Q(t) \langle\boldsymbol{\xi}\boldsymbol{\xi}^T\rangle Q(t)^T = Q(t)Q(t)^T = G(t)$,
with a proper choice of $Q(t)$.

However, to carry this out in practice encounters two challenges:
(i) given $G(t)$ the factor $Q(t)$ must be determined,
(ii) the matrix-vector multiplication $Q(t)\boldsymbol{\xi}$ must
be carried out.  For (i) the Cholesky algorithm 
is typically used with a computational cost
of $O(N^3)$, where $N$ is the number of components
of $\mathbf{g}$.  For (ii) the resulting factors 
$Q(t)$ are generally not sparse, which when
generating each variate incurs  a
computational cost of $O(N^2)$.  To get a sense of
the costs, for a three dimensional mesh, the number of components 
of $\mathbf{g}$ scales cubically as 
$N = (\ell/\Delta{x})^3$, where $\ell$ is the
domain size and $\Delta{x}$ is the mesh resolution.
The associated costs for generating the variates 
using this approach even for moderate spatial resolutions
is prohibitively expensive.  

To obtain a more efficient computational method we use
specific features of the discretization 
introduced in Section~\ref{sec_soft_materials_subject_shear}.
One useful feature of the discretization we use is that the equilibrium
covariance matrix is proportional to the identity matrix 
$C = \alpha \mathcal{I}$ with 
$\alpha = {2k_B{T}}/{3\rho_0 \Delta{x}^d}$.  
This allows equation~\ref{equ_time_G} to be expressed as
\begin{eqnarray}
\label{equ_deformFluctDissp_reduce}
G(t) = -2\alpha \wp(t) L(t).
\end{eqnarray}
We also use the following specific properties of
the operators $C$, $L(t)$, and $\wp(t)$ obtained from
the discretization.  The first is that each of the 
operators corresponds 
to use of numerical stencils which are 
translation invariant on the mesh.  This
has the important consequence that all of these 
operators are diagonalizable in the Fourier basis.
This has the further important consequence 
that all of these operators commute.  The second
is that $\wp$ is an exact projection operator,
so that $\wp^2 = \wp$ and $\wp = \wp^T$.  Finally, 
we use that the discrete approximation of the Laplacian 
is symmetric negative semi-definite so that it 
can be factored as $L(t) = -U(t)U^T(t)$ for some 
matrix $U(t)$.

By using these properties of the operators 
we can express the covariance of the stochastic driving
field as
\begin{eqnarray}
\label{equ_deformFluctDissp_2}
G(t) = \left( \sqrt{2\alpha} \wp U(t) \right) \left( \sqrt{2\alpha} \wp U(t) \right)^T.
\end{eqnarray}
In this form the required matrix square-root is readily obtained as 
$Q(t) = \sqrt{2\alpha} \wp U(t)$.  We remark this is different than 
the Cholesky factor obtained from $G(t)$ which is 
required to be lower triangular~\citep{Trefethen1997}.  Since the 
operators $L(t)$ and $\wp$ are diagonalizable in Fourier space, 
the matrix action of the
operators $U(t)$ and $\wp$ on any vector can be computed using the 
Fast Fourier Transform with a cost of $O(N \log(N))$.  In summary, our method allows 
in practice for the random variates of the stochastic driving field to 
be computed from $\mathbf{g} = Q(t)\boldsymbol{\xi}$ very efficiently,
with a computational cost of only $O(N \log(N))$.  This is in contrast to
the traditional Cholesky approach with a computational cost of $O(N^3)$.

\section{Regime II : Under-resolution of Hydrodynamic Relaxation (Quasi-Steady-State Limit)}
For many problems the equations of motion 
can be simplified by exploiting a separation of
time-scales between the time-scale on which the 
hydrodynamic modes relax to a statistical steady-state
and the time-scale associated with the motion of the 
microstructures.  In this case the fluid equations
can be approximated by 
\begin{eqnarray}
\label{equ_quasi_steady_stokes}
\mathbf{w} = -\tilde{L}(t)^{-1}\left[\Lambda + \mathbf{J}\right]  + \mathbf{a}.
\end{eqnarray}
The $\tilde{L} = \wp L \wp^T$ and the inverse is defined for the 
operator restricted to the linear space 
$\mathcal{V} = \{\mathbf{w} \in \mathbb{R}^{3N}|  \wp \mathbf{w} = \mathbf{w}\}$.  The term $\mathbf{a}$ is introduced 
to account for the thermal fluctuations in this regime.  We refer to this as the 
Quasi-Steady-State Stokes approximation~\citep{Brady1988}.  Using this 
in equation~\ref{equ_EL_L_discr}, we obtain the following closed 
system of equations for the motion of the microstructures
\begin{eqnarray}
\label{equ_X_closed_form}
\frac{d\mathbf{X}(t)}{dt} = H_{\mbox{\tiny SELM}}(t)\left[\mathbf{F}\right] + \bar{\mathbf{J}} + \mathbf{A}
\end{eqnarray}
where 
\begin{eqnarray}
\label{equ_H_el_def}
H_{\mbox{\tiny SELM}}(t) & = & -\Gamma \tilde{L}(t)^{-1} \Lambda\\
\bar{\mathbf{J}}         & = &  -\Gamma \tilde{L}(t)^{-1} \mathbf{J}.
\end{eqnarray}
The $\mathbf{A}$ will be used to account for the thermal fluctuations.  
We consider the specific case when the operator $\Lambda$ is linear
in $\mathbf{F}$ and $\Gamma$ is linear in $\mathbf{u}$.  In this case
$H_{\mbox{\tiny SELM}}$ is a tensor which we refer to as the 
''effective hydrodynamic coupling tensor.''

In this regime the thermal fluctuations arise from 
the hydrodynamic modes which are relaxed to statistical 
steady-state.  A key challenge is to determine the
appropriate statistics of $\mathbf{A}$ which accounts 
for the time integrated thermal fluctuations of the 
hydrodynamics which impact the microstructure dynamics.
For this purpose we rewrite equation~\ref{equ_X_closed_form}
in differential form as
\begin{eqnarray}
\label{equ_X_diff_form}
d\mathbf{X}(t) = H_{\mbox{\tiny SELM}}(t)\mathbf{F}dt + R(t)d\mathbf{B}_t
\end{eqnarray}
neglecting for the moment $\bar{\mathbf{J}}$, 
and representing the contributions of $\mathbf{A}$ by $R(t)d\mathbf{B}_t$.
We derive the covariance structure $S(t) = R(t)R(t)^T$ by requiring
consistency with the principle of Detailed-Balance of 
statistical mechanics~\citep{Reichl1998}.  The 
Fokker-Planck equation associated with equation 
\ref{equ_X_diff_form} is 
\begin{eqnarray}
\frac{\partial \Psi(\mathbf{X},t)}{\partial t}
& = & 
-\nabla \cdot \mathcal{J} \\
\mathcal{J} & = & 
H_{\mbox{\tiny SELM}}(t)\mathbf{F}\Psi
-\frac{1}{2}S(t)\nabla_{\mathbf{X}}\Psi.
\end{eqnarray}
The $\Psi(\mathbf{X},t)$ is the probability density for the 
microstructures to have configuration $\mathbf{X}$ at time $t$.
The equilibrium fluctuations of the system are required to 
have the Boltzmann distribution 
\begin{eqnarray}
\Psi_{BD}(\mathbf{X}) = \frac{1}{Z} \exp\left(-\Phi(\mathbf{X})/k_BT\right)
\end{eqnarray}
where $Z$ is a normalization constant which ensures the distribution
integrates to one~\citep{Reichl1998}.  
Substituting this above and using that $H_{\mbox{\tiny SELM}}$ is linear in 
$\mathbf{F}$ gives 
\begin{eqnarray}
\mathcal{J} & = & 
\left(
H_{\mbox{\tiny SELM}}(t)
-
\frac{1}{2 k_B T}S(t)
\right)\mathbf{F}\Psi_{BD}
\end{eqnarray}
where $\mathbf{F} = -\nabla_{\mathbf{X}}{\Phi}$.  The principle 
of Detailed-Balance requires at thermodynamic equilibrium that
$\mathcal{J} = 0$.   Requiring this to hold for all possible
$\mathbf{F}$ gives 
\begin{eqnarray}
\label{equ_def_S_DB}
S(t) = 2k_BT H_{\mbox{\tiny SELM}}(t).
\end{eqnarray}
For $S(t)$ to provide a covariance for a real-valued stochastic driving term, 
the hydrodynamic coupling tensor $H_{\mbox{\tiny SELM}}(t)$ 
must be symmetric and positive semi-definite.
In the case that $\Lambda$ and $\Gamma$ are linear operators this 
is ensured by condition~\ref{equ_coupling_cond_energy_discr},
which from expression~\ref{equ_H_el_def} gives
\begin{eqnarray}
\mathbf{q}^T H_{\mbox{\tiny SELM}}(t) \mathbf{q} = 
-\mathbf{v}^T\left(\tilde{L}(t)^{-1}\right) \mathbf{v}  \Delta{x}^d
\geq 0.
\end{eqnarray}
To obtain this result we let 
$\mathbf{v} = \Gamma^T\mathbf{q}$ and use that
$\tilde{L}(t)^{-1}$ is symmetric negative definite.
To obtain an approach useful in practice requires 
efficient methods for the generation of the stochastic 
driving term with covariance $S(t)$.

\subsection{Generating the Stochastic Driving Field II} 
\label{sec_gen_stoch_II}
As discussed in Section~\ref{sec_gen_stoch_I}, a significant challenge 
in practice is to generate efficiently the Gaussian 
stochastic driving terms with the required covariance structure.  
We discuss an approach for SELM methods when the coupling operators 
$\Lambda$ and $\Gamma$ are linear.  In this case
\begin{eqnarray}
H_{\mbox{\tiny SELM}}(t) & = & 
-\Gamma \tilde{L}(t)^{-1} \Gamma^T \Delta{x}^d
\end{eqnarray}
by condition~\ref{equ_coupling_cond_energy_discr}.
Using properties of the operators discussed in 
Section~\ref{sec_soft_materials_subject_shear}, 
we can express the hydrodynamic
coupling tensor as
\begin{eqnarray}
\label{equ_H_factor}
H_{\mbox{\tiny SELM}}(t) 
& = & \left(\Gamma(t) V(t) \Delta{x}^{d/2}\right)
\left(\Gamma(t) V(t) \Delta{x}^{d/2}\right)^T.
\end{eqnarray}
We have used that the operators 
$L(t)$ and $\wp$ commute and since $\wp$ is 
an exact projection that $\wp = \wp^T$, 
$\wp = \wp^2$.  Since $\tilde{L}(t)$ is symmetric
negative definite on the linear space 
$\mathcal{V} = \{\mathbf{w} \in \mathbb{R}^{3N}|  \wp \mathbf{w} = \mathbf{w}\}$,
we can factor $\tilde{L}(t)^{-1} = -V(t)V(t)^T$.
The factor $V(t)$ is readily obtained since $L(t)$
and $\wp$ are diagonalizable in the Fourier 
basis.  From equations~\ref{equ_H_factor} and~\ref{equ_def_S_DB} we 
can factor the covariance as $S(t) = R(t)R(t)^T$ with
\begin{eqnarray}
R(t) = \left(2 k_B{T}\Delta{x}^{d}\right)^{1/2} \Gamma(t) V(t).
\end{eqnarray}
This expression for the factor can be used to compute the required
Gaussian stochastic driving term $\mathbf{A} = R(t)\boldsymbol{\xi}$ 
with a computational cost of $O(N\log(N) + M)$, where $N$ is the total 
number of mesh sites in the momentum field discretization and assuming the 
action of $\Gamma$ can be computed with a cost of $O(M)$ with $M < N$.

The random variates are generated by utilizing 
the underlying discretization mesh
of the momentum equations.  This is accomplished 
by generating on the mesh uncorrelated standard 
Gaussian random variates $\boldsymbol{\xi}$.
Since $V(t)$ is diagonal in the Fourier basis, 
the action $V(t)\boldsymbol{\xi}$ is computed in 
Fourier space with a cost of only $O(N\log(N))$.
The operator $\Gamma$ is then applied.
If the operator $\Gamma(t)$ makes use of 
only localized values of the mesh it can be 
computed with computational cost of 
$O(M)$.  The last step in generating 
the random variate requires a scalar 
multiplication which
incurs a computational cost of 
$O(M)$.  This procedure generates the stochastic
driving term $\mathbf{A}$ with a computational 
cost of $O(N\log(N) + M)$.  For a sufficiently 
large number of microstructure degrees of freedom $M$,
this method is significantly more efficient than 
the traditional approach based on Cholesky 
factorization of $H_{\mbox{\tiny SELM}}$ which 
costs $O(M^3)$.

\subsubsection{Effective Hydrodynamic Coupling Tensor : $\mathbf{H}_{\mbox{\tiny SELM}}$}

We now discuss an approach for analyzing the effective hydrodynamic coupling 
tensors $\mathbf{H}_{\mbox{\tiny SELM}}$ which appear in the quasi-steady-state
formulation of the SELM approach.  From equation~\ref{equ_H_el_def} many 
types of hydrodynamic coupling tensors are possible depending on the kinetic
constraints and choice of coupling operators $\Lambda$ and $\Gamma$.  For 
concreteness we discuss the specific case corresponding to the 
Stochastic Immersed Boundary Method (SIB)~\citep{Peskin2002,Atzberger2007a}.  
In the case of the SIB method, the specific coupling operators $\Lambda$ and 
$\Gamma$ are given by~\ref{equ_op_SIB_Lambda} 
and~\ref{equ_op_SIB_Gamma}.  From equation~\ref{equ_H_el_def} the effective hydrodynamic 
coupling tensor is given by 
\begin{eqnarray}
\label{equ_def_H_IB}
\\
\nonumber
\left[\mathbf{H}_{\mbox{\tiny IB}}(t)\left[\mathbf{F}\right]\right]^{[j]} & = & 
-\sum_{\mathbf{m}} \delta_a(\mathbf{x}_{\mathbf{m}} - \mathbf{X}^{[j]}(t))
\left[
\tilde{L}(t)^{-1}\left(\sum_{j = 1}^{M} \mathbf{F}^{[j]}
\delta_{a}(\mathbf{x}_{\mathbf{m}} - \mathbf{X}^{[j]}(t))\right) \right]_{\mathbf{m}}\Delta{x}^d.
\end{eqnarray}
In the notation, the superscript $[\cdot]^{[j]}$ denotes for the composite vector the components 
associated with the $j^{th}$ microstructure degree of freedom.  The $[\cdot]_{\mathbf{m}}$
denotes the vector components associated with the mesh site with index $\mathbf{m}$.
An analysis of variants of this tensor for point particles and slender bodies was 
carried-out in~\citep{Bringley2008,Atzberger2007c}.  

Since $\mathbf{H}_{\mbox{\tiny IB}}$ is linear in the microstructure forces, 
without loss of generality we can consider the case of only 
two microstructure degrees of freedom.  We denote these as 
$\mathbf{X}^{[1]}$, $\mathbf{X}^{[2]}$ and the displacement
vector by $\mathbf{z} = \mathbf{X}^{[2]} - \mathbf{X}^{[1]}$.
In making comparisons with other hydrodynamic coupling tensors
we find it helpful to make use of approximate symmetries
satisfied by $\mathbf{H}_{\mbox{\tiny IB}}$.  From equation~\ref{equ_def_H_IB}, 
$\mathbf{H}_{\mbox{\tiny IB}}$ depends on $\mathbf{z}$ up to a shift of 
$\mathbf{X}^{[1]}$ relative to the nearest mesh site, and is 
similarly rotationally symmetry about the axis of $\mathbf{z}$.
This allows the tensor components for all configurations to
be related to a canonical configuration with 
$\mathbf{z} = (z_1, 0, 0)$.  For any configuration
this is accomplished by introducing the rotation 
matrix $U$ so that $U\mathbf{z} = (z_1, 0, 0)$ and 
considering $\tilde{\mathbf{H}} = U\mathbf{H}_{\mbox{\tiny IB}}U^T$.  
In our comparisons we consider 
$\bar{\mathbf{H}}_{\mbox{\tiny IB}} = \langle \tilde{\mathbf{H}} \rangle$, where the average 
is taken over all rotations and shifts 
with respect to the nearest mesh site.

\begin{figure}[t*]
\centering
\epsfxsize = 5in
\epsffile[0 0 360 231]{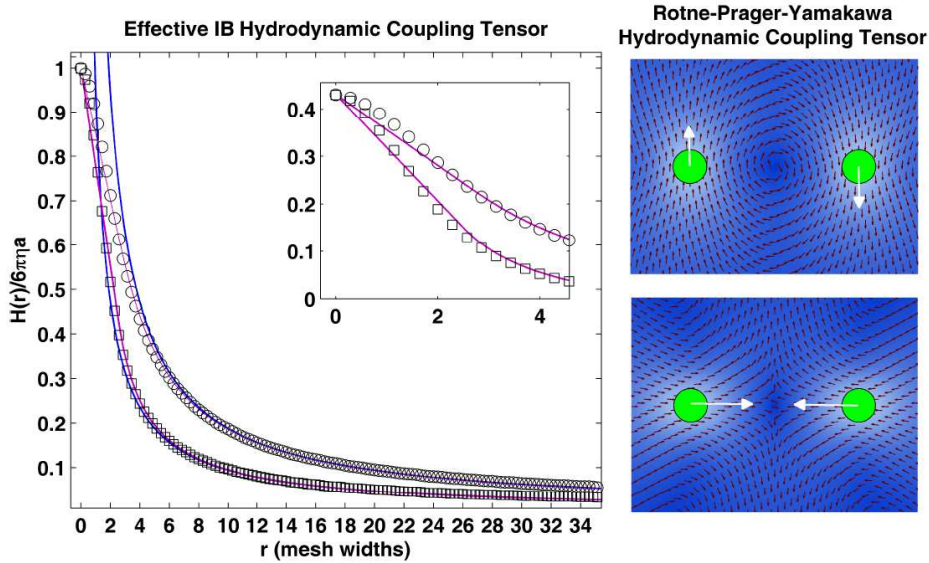}
\caption[Effective Hydrodynamic Coupling Tensors]
{Comparison of the Hydrodynamic Coupling Tensor of the 
Immersed Boundary Method 
$\mathbf{H}_{\mbox{\tiny IB}}$ with the Oseen Tensor
$\mathbf{H}_{\mbox{\tiny OS}}$ 
and the Rotne-Prager-Yamakawa Tensor $\mathbf{H}_{\mbox{\tiny RPY}}$.  The components of 
the hydrodynamic coupling tensor for displacement $r = |\mathbf{z}|$ are shown 
for the parallel direction (circles) and the perpendicular direction (squares).
For two particles subject to an equal and opposite force, the velocity field 
corresponding to the Rotne-Prager-Yamakawa Tensor is shown on the right.  
The lighter shaded regions indicate a larger magnitude of the velocity.  
%In both plots the computations used an effective particle size of 
%$a = 4.70\mbox{nm} = 1.25\mbox{mesh-widths}$.
}
\label{figure_compare_IB_RPY_OS}
\end{figure}

In practice, to numerically compute
$\bar{\mathbf{H}}_{\mbox{\tiny IB}}$
we sample random configurations of
$\mathbf{X}^{[1]}$ and $\mathbf{X}^{[2]}$.
A useful expression for the tensor components
is $H_{ij} = \mathbf{e}_i^T \mathbf{H} \mathbf{e}_j = \mathbf{e}_i^T \mathbf{v}$.
In this notation, $\mathbf{e}_k$ are the standard basis vectors in direction $k$
and $\mathbf{v}$ is the microstructure velocity.  For a computational implementation
of the SELM method, this can be 
used by applying the force $\mathbf{e}_j$
to the microstructure degrees of freedom and measuring the components of the
realized microstructure velocities $\mathbf{v}$.

When using a SELM approach the hydrodynamic 
coupling tensor has features which depend on the 
discretization of the momentum equations,
discretization of the microstructures, and 
the specific choice of coupling operators.
For the specific choice of the IB coupling 
operators and discretization on a uniform mesh 
we discuss how the effective hydrodynamic 
coupling tensor compares with other hydrodynamic
coupling tensors.  We consider two specific 
tensors, the Oseen Tensor~\citep{Brady1988} 
and the Rotne-Prager-Yamakawa 
Tensor~\citep{Rotne1969,Yamakawa1970}.
The Oseen Tensor for a pair of particles 
experiencing equal and opposite forces 
can be expressed in terms of the displacement vector
$\mathbf{z}$ as
\begin{eqnarray*}
\mathbf{H}_{\mbox{\tiny \textbf{OS}}}(\mathbf{z}) 
& = & 
\frac{2}{6\pi\eta a}\left[
\mathcal{I} - \frac{3}{4}
\frac{a}{r}
\left(
\mathcal{I}
+ 
\frac{\mathbf{z}\mathbf{z}^T}{r^2}
\right)
\right].
\end{eqnarray*}
Similarly, the Rotne-Prager-Yamakawa 
Tensor can be expressed in terms of the displacement vector
$\mathbf{z}$ as
\begin{eqnarray*}
\mathbf{H}_{\mbox{\tiny \textbf{RPY}}}(\mathbf{z}) 
& = & 
\frac{2}{6 \pi \eta a}
\left[
\mathcal{I}
-
\frac{3}{4} \frac{a}{r}
\left\lbrace
\begin{array}{ll}
\left(1 + \frac{2a^2}{3r^2}\right)\mathcal{I} 
+ 
\left(1 - \frac{2a^2}{r^2}\right)
\frac{\mathbf{z}\mathbf{z}^T}{r^2},
&\mbox{\small for $r \geq 2a$} \\
\frac{r}{2a}
\left[
\left(\frac{8}{3} - \frac{3r}{4a}\right)\mathcal{I} 
+ 
\frac{r}{4a}\frac{\mathbf{z}\mathbf{z}^T}{r^2}
\right],
&\mbox{\small for $r < 2a$} \\
\end{array}
\right\}
\right]
.
\end{eqnarray*}
In the notation, $\eta$ denotes the dynamic fluid viscosity,
and $a$ denotes the effective particle size in terms of 
the radius of a sphere.

In Figure~\ref{figure_compare_IB_RPY_OS} the $\mathbf{H}_{\mbox{\tiny IB}}$
is compared with the Oseen Tensor $\mathbf{H}_{\mbox{\tiny OS}}$ and 
Rotne-Prager-Yamakawa Tensor $\mathbf{H}_{\mbox{\tiny RPY}}$.  It is 
found that the effective hydrodynamic coupling tensor of the Immersed Boundary 
Method agrees well with both of the tensors in the far-field $r \gg a$.  
An interesting finding is that in the near-field $\mathbf{H}_{\mbox{\tiny IB}}$
shows very close agreement to $\mathbf{H}_{\mbox{\tiny RPY}}$, see inset in 
Figure~\ref{figure_compare_IB_RPY_OS}.  

\section{Applications}
The SELM approach is expected to be applicable in the study of many
different types of complex fluids and soft materials.  As a demonstration
of the proposed stochastic numerical methods, simulation studies are 
carried out for a few specific systems.  These include studying:
(i) the dependence of the shear viscosity on the shear rate
in a FENE polymeric fluid,
(ii) the frequency response of the elastic storage modulus and viscous loss
modulus of a lipid vesicle fluid subject to oscillatory shear, (iii) 
the rheological responses over time of a gel-like material subject 
to a constant rate of shear.  We now discuss each of these 
simulation studies in detail.

\subsection{Estimating Effective Macroscopic Stress}
\label{sec_stress_estimator}
An important challenge in the study of complex fluids and soft materials
is to relate bulk material properties to phenomena on the level of 
the microstructures of the material.  To characterize properties
of a material, experimental measurements are often made as a
sample of material is subject to shear~\citep{Bird1987Vol1,Bird1987Vol2}.  
To link 
microstructure mechanics, interactions, and kinetics to 
macroscopic material properties we develop estimators for
an effective macroscopic stress tensor.  Our 
estimators are based on similar approaches used to obtain
the Irving-Kirkwood-Kramer formulas~\citep{Irving1950,Doi1986,Bird1987Vol1,Bird1987Vol2}.

When using the SELM approach, the microstructures are modeled using 
n-body interactions and the domain is subject to generalized boundary 
conditions.  For example, 
two body interactions can 
arise from bonds between monomer particles and three body interactions
can arise from bond angle terms included in the potential energy.
Estimators for the stress must take these features into account.

\begin{figure}[t*]
\centering
\epsfxsize = 5in
\epsffile[14 14 375 153]{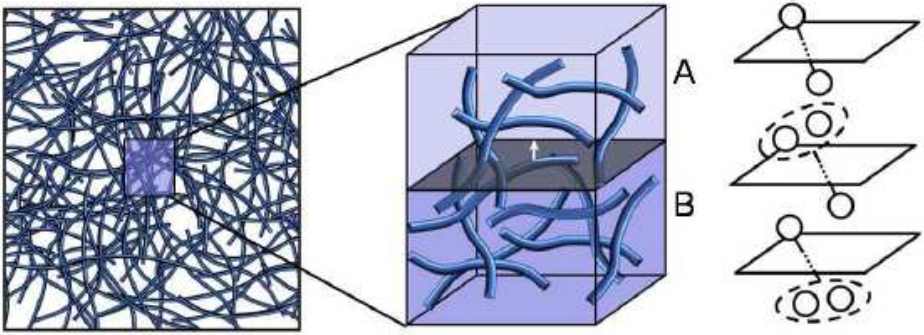}
\caption[Macroscopic Stress Estimator]
{An effective macroscopic stress is estimated from a sample of the 
material by computing the forces transmitted across a plane 
which cuts through the sample at a specified location and 
with a specified normal.  At the level of the microstructures,
the cut-plane is used to divide the 
sample into two bodies labeled $A$ and $B$, shown in the middle.  
The effective stress is estimated by computing the force exerted 
by particles in body $A$ on particles in body $B$.  For models with two body 
interactions a contribution is made to the stress only if 
one particle is in body $A$ while the other is in body $B$,
shown on the far right on top.
For three body interactions there are two possible cases for
how forces can be transmitted across the cut plane, shown
on the far right in the middle and bottom.
}
\label{figure_stressEstimator}
\end{figure}

To obtain a notion of macroscopic stress we define a normal direction and 
a plane which cuts the unit cell.  We then determine on average the forces 
exerted by the particles which lie above this plane on the particles which lie 
below this plane.  We define the effective stress associated with this plane 
as the total of this exerted force divided by the area of the plane.  To define 
an effective macroscopic stress we average over all possible planes within the 
unit cell having the specified normal direction, see Figure~\ref{figure_stressEstimator}.

More precisely, the effective macroscopic stress arising from 
n-body interactions is estimated using 
\begin{eqnarray}
\label{equ_init_stress_estimator}
\sigma_{\ell,z}^{(n)} = \frac{1}{L} \left \langle \int_a^b \Lambda_{\ell,z}^{(n)}(\zeta) d\zeta \right \rangle.
\end{eqnarray}
The $L = b - a$ is the length of the domain in the $z$-direction
and
$<\cdot>$ denotes averaging over the ensemble.
The $\Lambda_{\ell,z}^{(n)}$ denotes the microscopic 
stress arising from the $n$-body interactions associated 
with a given stress plane and is defined by
\begin{eqnarray}
\label{equ_micro_stress_estimator}
\Lambda_{\ell,z}^{(n)}(\zeta) 
& = & 
\frac{1}{A} 
\sum_{\mathbf{q} \in \mathcal{Q}_n}
\sum_{k = 1}^{n - 1}
\sum_{j = 1}^{k}
\mathbf{f}_{\mathbf{q}, j}^{(\ell)}
\prod_{j = 1}^{k}
\mathcal{H}(\zeta - \mathbf{x}_{q_j}^{(z)})
\prod_{j = k + 1}^{n}
\mathcal{H}(\mathbf{x}_{q_j}^{(z)} - \zeta).
\end{eqnarray}
The $\mathcal{Q}_{n}$ is the set of $n$-tuple 
indices $\mathbf{q} = (q_1,\ldots,q_n)$
describing the $n$-body interactions of the system,
$\mathbf{f}_{\mathbf{q}, j}$ denotes the force
acting on the $j^{th}$ particle of the interaction,
and $\mathbf{x}_{q_j}$ denotes the $j^{th}$ 
particle involved in the interaction.  As a matter 
of convention in the indexing $\mathbf{q}$
we require that
$i \leq j$ implies $\mathbf{x}_{q_i}^{(z)} 
\leq \mathbf{x}_{q_j}^{(z)}$.  This 
expression corresponds to a sum over all 
the forces exerted by particles of the material
above the cross-section at $\zeta = z$ on the 
particles of the material below.  Each term
of the summation over $k = 1, \ldots, n - 1$
corresponds to a specific number of particles 
of the $n$-body interaction lying below the 
cross-section at $\zeta = z$, see Figure~\ref{figure_stressEstimator}.

When integrating the microscopic stress, a useful identity is that
\begin{eqnarray}
\int_a^b \Pi_{j = 1}^{k}
\mathcal{H}(\zeta - \mathbf{x}_{q_j}^{(z)})
\cdot
\Pi_{j = k + 1}^{n}
\mathcal{H}(\mathbf{x}_{q_j}^{(z)} - \zeta) 
d\zeta = \mathbf{x}_{q_{k + 1}}^{*,(z)} - \mathbf{x}_{q_k}^{*,(z)}
\end{eqnarray}
where 
\begin{eqnarray}
\mathbf{x}_{q_j}^{*,(z)} = 
\left\{
\begin{array}{ll}
b,                                & \mbox{if $\mathbf{x}_{q_j}^{(z)} \geq b$} \\
\mathbf{x}_{q_j}^{(z)},           & \mbox{if $a \leq \mathbf{x}_{q_j}^{(z)} \leq b$} \\
a,                                & \mbox{if $\mathbf{x}_{q_j}^{(z)} \leq a$}. \\
\end{array}
\right.
\end{eqnarray}
By integrating equation~\ref{equ_micro_stress_estimator} we obtain
\begin{eqnarray}
\int_a^b
\Lambda_{(\ell),z}^{(n)}(\zeta) 
d\zeta
= \frac{1}{A} 
\sum_{\mathbf{q} \in \mathcal{Q}_n}
\sum_{k = 1}^{n - 1}
\sum_{j = 1}^{k}
\mathbf{f}_{\mathbf{q}, j}^{(\ell)}
\cdot
\left(
\mathbf{x}_{q_{k + 1}}^{*,(z)} - \mathbf{x}_{q_k}^{*,(z)}
\right).
\end{eqnarray}
This can be further simplified by switching the order of summation of $j$ and $k$
and using the telescoping property of the summation over $k$.  From 
equation~\ref{equ_micro_stress_estimator} this yields the following 
estimator for the stress contributions of the $n$-body interactions
\begin{eqnarray}
\label{equ_stress_estimator}
\sigma_{\ell,z}^{(n)} = \frac{1}{AL} 
\sum_{\mathbf{q} \in \mathcal{Q}_n}
\sum_{j = 1}^{n - 1}
\left \langle 
\mathbf{f}_{\mathbf{q}, j}^{(\ell)}
\cdot
\left(
\mathbf{x}_{q_n}^{*,(z)} - \mathbf{x}_{q_j}^{*,(z)}
\right)
\right \rangle.
\end{eqnarray}
This defines an effective macroscopic stress tensor contribution
in terms of the n-body interactions of the microstructures of the material.
To obtain the total contribution of the microstructure interactions to
the stress, all of the contributions of the n-body interactions are summed
to obtain the effective macroscopic stress tensor
\begin{eqnarray}
\sigma_{\ell,z} = \sum_n \sigma_{\ell,z}^{(n)}.
\end{eqnarray}
This notion of the macroscopic stress will be used to link bulk 
rheological properties to the microscopic simulations.

\subsection{Application I: Complex Fluid of Finite Extensible Non-linear Elastic (FENE) Dimers}

\begin{figure}[t*]
\centering
\epsfxsize = 5.5in
\epsffile[-218   181   830   610]{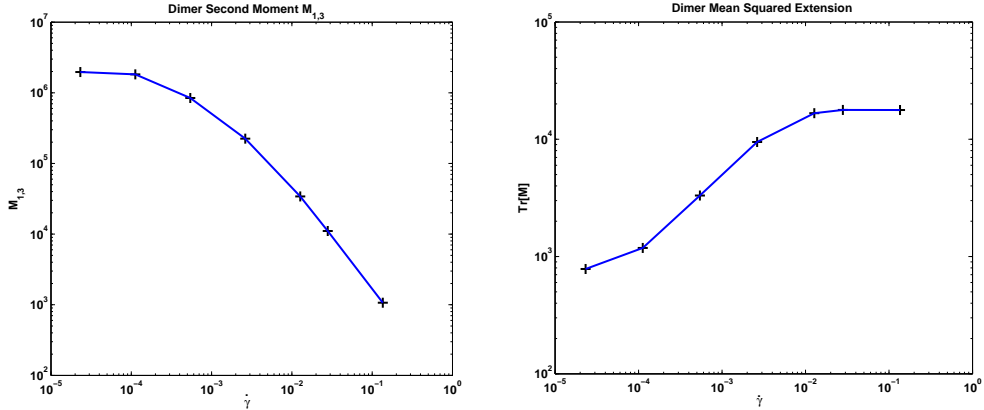}
%\epsffile[14 14 375 174]{fig_FENE_moments2.eps}
\caption[Moments of Extension Vector in FENE Fluid Simulations]
{Components of the second moment of the extension vector are shown as the shear
rate is varied.  The second moment matrix is $M = \langle \mathbf{z} \mathbf{z}^T\rangle$.
On the left is shown the off diagonal entry $M_{1,3}$ as a function of shear rate. 
On the right is shown the averaged mean squared extension vector of the dimer,
which is given by $\bar{\ell}^2 = \langle |\mathbf{z}|^2\rangle = \mbox{Trace}[M]$.
The moments show a significant dependence on the rate of shear.
}
\label{figure_compare_FENE_moments}
\end{figure}

As a demonstration of the proposed computational methodology we  
consider a fluid with microstructures consisting of elastic polymers.
The polymers are modeled as idealized elastic dimers which have the potential 
energy
\begin{eqnarray}
\label{equ_V_r}
\phi(r) = \frac{1}{2} K r_0^2 \log\left(1 - \left(\frac{r}{r_0}\right)^2\right).
\end{eqnarray}
The $K$ denotes the polymer stiffness, $r$ denotes the length of extension of the 
dimer, and $r_0$ denotes the maximum permitted extension length~\citep{Bird1987Vol2}.  
The configuration of each dimer will be represented using 
two degrees of freedom $\mathbf{X}^{(1)}$, $\mathbf{X}^{(2)}$.
The potential energy for the dimer is given by
$\Phi(\mathbf{X}) = \phi(|\mathbf{X}^{(2)} - \mathbf{X}^{(1)}|)$,
where $\mathbf{X}$ is the composite vector for the particle configuration.

\begin{table}[t]
\centering
\begin{tabular}{|l|l|}
\hline
Parameter & Description \\
\hline
$N$                           & Number of mesh points in each direction.\\
$\Delta{x}$                   & Mesh spacing.                  \\
$L$                           & Domain size in each direction. \\
$T$                           & Temperature.                   \\
$k_B$                         & Boltzmann's constant. \\                              
$\mu$                         & Dynamic viscosity of the solvent fluid. \\                             
$\rho$                        & Mass density of the solvent fluid. \\                             
$K$                           & Bond stiffness. \\
$r_0$                         & Maximum permissible bond extension. \\
$\gamma_s$                    & Stokesian drag of a particle. \\
$\dot{\gamma}^0$              & Shear rate amplitude. \\
$\gamma^0$                    & Strain rate amplitude. \\
$a$                           & Effective radius of particle estimated via Stokes drag. \\
\hline
\end{tabular}
\caption[FENE Parameter Description]
{Description of the parameters used in simulations of the FENE polymeric fluid.
\label{table_FENE_param_descr}}
\end{table}

\begin{table}[t]
\centering
\begin{tabular}{|l|l|}
\hline
Parameter & Value \\
\hline   
$N$                           & 36 \\
$\Delta{x}$                   & $11.25 \mbox{ nm}$ \\
$L$                           & $405 \mbox{ nm}$ \\
$T$                           & $300 \mbox{ K}$ \\
$k_B$                         & $8.3145 \times10^3    \mbox{ nm}^2\cdot\mbox{amu}\cdot\mbox{ns}^{-2}\cdot\mbox{K}^{-1}$ \\
$\mu$                         & $6.0221 \times 10^{5} \mbox{ amu}\cdot\mbox{cm}^{-1}\cdot\mbox{ns}^{-1}$ \\
$\rho$                        & $6.0221 \times 10^{2} \mbox{ amu}\cdot\mbox{nm}^{-3}$ \\
$K$                           & $8.9796 \times 10^3 \mbox{ amu}\cdot\mbox{ns}^{-2}$ \\
$r_0$                         & $200 \mbox{ nm}$ \\
$\gamma_s$                    & $1.7027 \times 10^8 \mbox{ amu}\cdot\mbox{ns}^{-1}$  \\
$a$                           & $15 \mbox{ nm}$ \\
\hline
\end{tabular}
\caption[FENE Parameter Values]
{Values of the parameters used in simulations of the FENE polymeric fluid.
\label{table_FENE_value_descr}}
\end{table}

When the polymeric fluid is subject to shear 
the thermally fluctuating polymeric microstructures 
are expected to significantly re-orient and deform 
as a consequence of the shear stresses.  This along
with thermal fluctuations of the microstructures is 
expected to play an important role in the bulk
response of the polymeric fluid.  To link the bulk 
material properties of the fluid to the microstructures, 
we use the effective macroscopic 
stress $\sigma_p$ obtained from equation~\ref{equ_stress_estimator}.  To characterize 
the bulk rheological response we consider the shear viscosity $\eta_p$ 
and first normal stress coefficient $\Psi_1$ of the polymeric
fluid.  We define these as~\citep{Bird1987Vol1,Bird1987Vol2}
\begin{eqnarray}
\label{equ_eta_p_def_FENE}
\eta_p & = & {\sigma_p^{(s,v)}}/{\dot{\gamma}} \\
\Psi_1 & = & ({\sigma_p^{(s,s)} - \sigma^{(v,v)}})/{\dot{\gamma}^2}.
\end{eqnarray}
The $\dot{\gamma}$ is the rate of shear of the polymeric fluid.
In the notation, the superscript $(s,v)$ indicates the 
tensor component with the index $s$ corresponding to the 
direction of shear and the index $v$ corresponding to 
the direction of the fluid velocity.  The contributions of the 
solvent fluid to the shear viscosity and normal stresses can be 
considered separately~\citep{Bird1987Vol2}.

\begin{figure}[t*]
\centering
\epsfxsize = 5.5in
%\epsffile[-218   181   830   610]{fig_FENE_shearViscosity_firstNormalStressDiff.eps}
\epsffile[14 14 375 168]{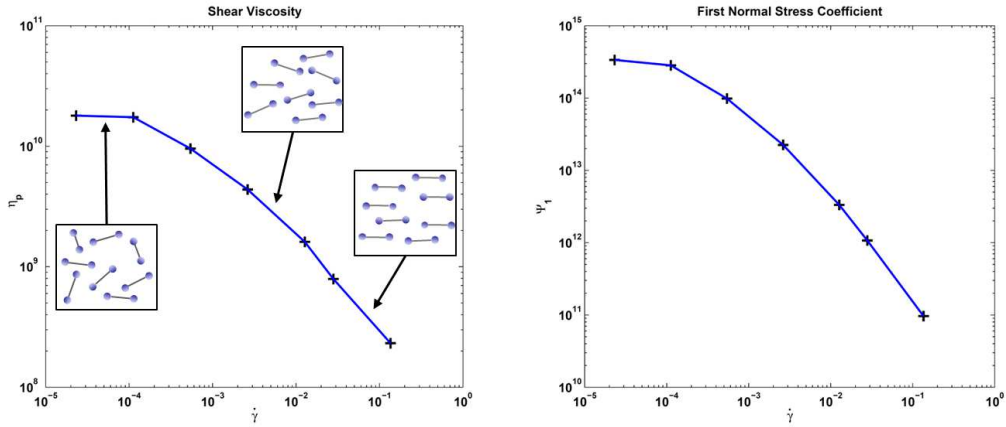}
\caption[Rheological Properties of FENE Polymeric Fluid]
{Rheological properties of the FENE polymeric fluid are shown as the rate of shear is varied.
The shear viscosity is shown on the left and the first normal stress difference 
is shown on the right.  As the shear rate increases the dimers align increasingly with 
the direction of fluid flow, shown as insets.
}
\label{figure_FENE_rheology}
\end{figure}

The SELM approach is used to study how the shear viscosity 
and first normal stress difference depend on the rate 
of shear of the polymeric fluid.  Simulations are performed 
using the SELM method in the regime where the hydrodynamic 
modes are relaxed to statistical steady-state with 
parameters given in Table~\ref{table_FENE_param_descr}.  For 
$\Lambda$ and $\Gamma$ the coupling tensors of 
equation~\ref{equ_op_SIB_Lambda} and equation~\ref{equ_op_SIB_Gamma} 
are used.  From an ensemble average over many computational
experiments the moments of the extension vector
$\mathbf{z}$ are estimated as the shear rate is increased.
The polymeric microstructure moments are seen to respond 
strongly as the shear stresses of the fluid increase, 
see Figure~\ref{figure_compare_FENE_moments}.  This
indicates that the rheological properties of the polymeric
fluid will depend significantly on the rate of shear.  
The SELM simulations show that the shear viscosity 
and the first normal stress difference do in fact 
vary significantly with the shear rate, see 
Figure~\ref{figure_FENE_rheology}.   

The shear viscosity is found to decrease as the shear rate 
increases.  This appears to occur as a consequence of the dimers 
increasingly aligning with the direction of the fluid flow and 
as a consequence of the dimers approaching the maximal extension 
permitted by equation~\ref{equ_V_r}.  The increased extension 
results in a non-linear increase in the effective stiffness 
of the dimer (defined for a given extension by Taylor expanding to
second order equation~\ref{equ_V_r}).  While the dimers become 
increasingly extended with stronger restoring forces
this is counter-balanced by the dimers being increasingly stiff
and the thermal fluctuations less frequently driving the 
dimer into configurations crossing the stress plane.  The net effect 
is that the mechanical stress transmitted on average by the dimers
in the direction of shear does not increase as the shear rate increases.  
This results in a lower effective shear viscosity (note the division by 
$\dot{\gamma}$ in equation~\ref{equ_eta_p_def_FENE}).  This is a 
well-known phenomena in polymeric fluids and is referred to as 
shear thinning.  The simulations demonstrate that the SELM approach 
is capable of capturing at the level of the microstructures such 
phenomena, see Figure~\ref{figure_FENE_rheology}.

\subsection{Application II: Polymerized Lipid Vesicle Fluid}

As a further demonstration of the applicability of the SELM
approach we show how the stochastic numerical methods can be 
used to investigate the bulk material properties of a complex fluid 
with polymerized vesicle microstructures.  We discuss how the 
methods can be used to compute the response of the complex 
fluid subject to an oscillating shear flow varied over a 
wide range of frequencies.

\begin{figure}[t*]
\centering
\epsfxsize = 5in
\epsffile[14 14 375 112]{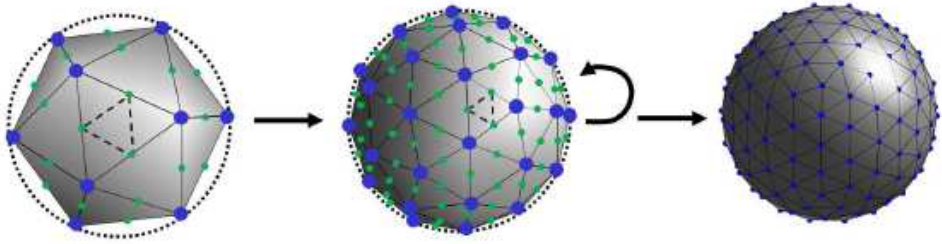}
\caption[Vesicle Mesh Construction]
{Recursive Method for Mesh Construction.  The triangulated
mesh for a spherical vesicle is constructed by starting with 
the vertices and faces of a regular icosahedron, shown on the 
left.  The edges of the icosahedron are bisected and connected
to divide each triangular face into four smaller triangular 
faces.  The vertices located at the bisection points are 
projected radially outward to the surface of the sphere,
shown in the middle.  This refinement procedure is repeated 
recursively until a mesh of sufficient resolution is obtained.
The mesh obtained after two levels of recursive refinement,
which we use to represent polymerized vesicles,
is shown on the right.
}
\label{figure_vesicles_meshAlg}
\end{figure}

To obtain a triangulated mesh which captures the shape of 
a vesicle having a spherical geometry we start with an 
icosahedral which is circumscribed by a sphere of a given 
radius.  We use the faces of the icosahedron as an initial
triangulated mesh.  To obtain a mesh which better approximates 
the sphere we bisect the three edges of each triangular face 
to obtain four sub-triangles.  The newly introduced vertices 
are projected radially outward to the surface of the sphere.  
The process is then repeated recursively to obtain further 
refinements of the mesh.  This yields a high quality mesh 
for spherical geometries.  A vesicle represented by a mesh
obtained using two levels of recursive refinement is 
shown in Figure~\ref{figure_vesicles_meshAlg}.

To account for the mechanics of a polymerized vesicle 
the following interactions are used for the control points 
of the mesh
\begin{eqnarray}
\phi_1(r,\ell) & = & 
\frac{1}{2}K_1\left(r - \ell\right)^2 \\
\phi_2(\boldsymbol{\tau}_1,\boldsymbol{\tau}_2) & = & 
\frac{1}{2}K_1 \left|\boldsymbol{\tau}_{1} - \boldsymbol{\tau}_{2}\right|^2.
\end{eqnarray}
The $r$ denotes the displacement between two control points, $\ell$ denotes 
a preferred distance between control points, and $\boldsymbol{\tau}$ denotes
a normalized displacement vector (tangent vector) between two control points.
The $\phi_1$ energy accounts for the stretching of a bond between two control
points beyond its preferred extension.  The $\phi_2$ energy accounts for 
bending of the surface locally by penalizing the misalignment of tangent
vectors.

For a given triangulated mesh of control points the total energy is given by
\begin{eqnarray}
\Phi[\mathbf{X}] & = & E_1[\mathbf{X}] + E_2[\mathbf{X}] \\
E_1[\mathbf{X}] & = & 
\sum_{(i,j) \in \mathcal{Q}_1} \phi_1(r_{ij},\ell_{ij}) \\
E_2[\mathbf{X}] & = & 
\sum_{(i,j,k) \in \mathcal{Q}_2} \phi_2(\boldsymbol{\tau}_{ij},\boldsymbol{\tau}_{jk}).
\end{eqnarray}
The $\mathbf{X}$ denotes the composite vector of control points.  The 
$j^{th}$ control point is denoted by $\mathbf{X}^{[j]}$.  The $\mathcal{Q}_1$
and $\mathcal{Q}_2$ are index sets defined by the topology of the triangulated 
mesh.

The first energy term $E_1$ accounts for stretching of the vesicle surface and is 
computed by summing over all local two body interactions $\mathcal{Q}_1$
defined by the topology of the triangulated mesh.  For the distance 
$r_{ij} = |\mathbf{X}^{[i]} - \mathbf{X}^{[j]}|$ between the two points 
having index $i$ and $j$, the energy $E_1$ penalizes deviations from
the preferred distance $\ell_{ij}$.
The preferred distances $\ell_{ij}$ are defined by the geometry 
of a spherical reference configuration for the vesicle.  To ensure the
two body interactions are represented by a unique index in $\mathcal{Q}_1$ 
we adopt the convention that $i < j$.

The second energy term $E_2$ accounts for curvature of the vesicle surface and 
is computed by summing over all local three body interactions $\mathcal{Q}_2$ 
defined by the topology of the triangulated mesh.  The energy penalizes the 
the misalignment of the tangent vectors 
$\boldsymbol{\tau}_{ij} = (\mathbf{X}^{[i]} - \mathbf{X}^{[j]})/r_{ij}$ and
$\boldsymbol{\tau}_{jk} = (\mathbf{X}^{[j]} - \mathbf{X}^{[k]})/r_{jk}$. 
In the set of indices in $\mathcal{Q}_2$ it is assumed
that the point with index $j$ is always adjacent to both 
$i$ and $k$.  To ensure the three body interactions are represented by a unique 
index in $\mathcal{Q}_2$ we adopt the convention that $i < k$.

\begin{figure}[t*]
\centering
\epsfxsize = 4in
\epsffile[14 14 375 253]{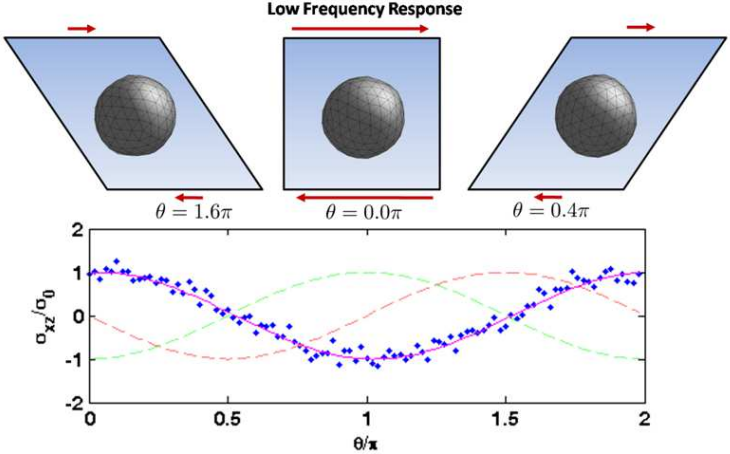}
%\epsfxsize = 9.4in
%\epsffile[0 0 660 120]{fig_vesicle_sheared_select_fits2.eps}
\caption[Vesicle Response]
{Simulation results showing the vesicle response when subject 
to an oscillating shear flow.
At low frequency the vesicle shape distortion is small 
and is masked by thermal fluctuations.  At low frequency 
the vesicle membrane stresses equilibrate 
to a good approximation with the bulk shear stresses,
as illustrated in the plot of $\sigma_{xz}$ .  
For the vesicle configurations 
shown, the low frequency response corresponds to 
$\omega =  3.9294 \times 10^{-3} \mbox{ns}^{-1}$,
$\dot{\gamma} = 1.9647 \times 10^{-3} \mbox{ns}^{-1}$,
$\sigma_0 = 3.7114\times 10^{8} \mbox{amu}\cdot\mbox{nm}^{-1}\cdot\mbox{ns}^{-2}$.
The phase $\theta = \omega{t}$ is reported in the range 
$[0,2\pi)$.  For additional parameters used in the simulations
see Table~\ref{table_vesicle_param_descr} and~\ref{table_vesicle_param_value}.}
\label{figure_vesicles_sheared_low_freq}
\end{figure}

\begin{figure}[t*]
\centering
\epsfxsize = 4in
\epsffile[14 14 375 253]{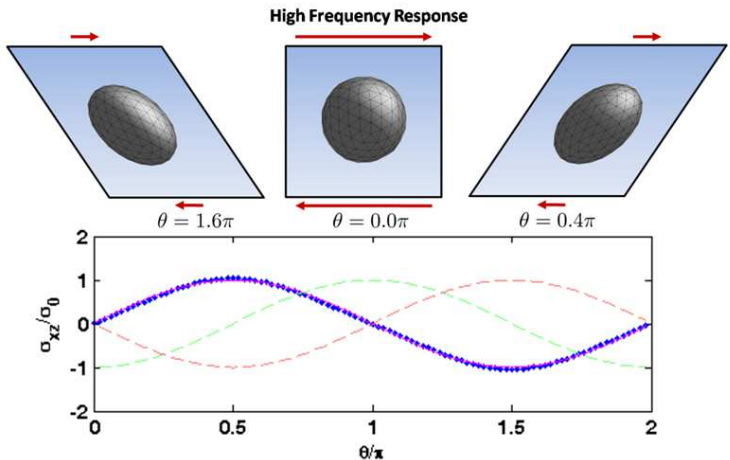}
%\epsfxsize = 9.4in
%\epsffile[0 0 660 120]{fig_vesicle_sheared_select_fits2.eps}
\caption[Vesicle Response]
{Simulations results showing the vesicle response when subject 
to an oscillating shear flow.  
At high frequency 
the vesicle shape is visibly distorted and the membrane 
stresses do not have time to equilibrate with the bulk
shear stresses, as illustrated by the configurations 
for phase $\theta = 1.6, 0.4$ and the plot of $\sigma_{xz}$.
For the vesicle configurations 
shown, the 
high frequency response corresponds to 
$\omega = 1.2426 \times 10^2 \mbox{ns}^{-1}$,
$\dot{\gamma} = 6.2129\times 10^{1} \mbox{ns}^{-1}$,
$\sigma_0 = 4.6314\times 10^{10} \mbox{amu}\cdot\mbox{nm}^{-1}\cdot\mbox{ns}^{-2}$.
The phase $\theta = \omega{t}$ is reported in the range 
$[0,2\pi)$.  For additional parameters used in the simulations
see Table~\ref{table_vesicle_param_descr} and~\ref{table_vesicle_param_value}.}
\label{figure_vesicles_sheared_high_freq}
\end{figure}

To investigate the bulk rheological properties, the complex 
vesicle fluid is subjected to an oscillatory shear with 
rate $\dot{\gamma} = \dot{\gamma}^0\cos(\omega t)$.
We consider the dilute regime in which it is sufficient
to study a single polymerized vesicle subject to 
oscillatory shear.  To estimate the effective macroscopic stress 
tensor the tensor is decomposed into contributions from 
two body and three body interactions 
\begin{eqnarray}
\sigma_{\ell,z} = \sigma_{\ell,z}^{(2)} + \sigma_{\ell,z}^{(3)}.
\end{eqnarray}
For the contributions of the n-body interactions to 
the macroscopic stress $\sigma_{\ell,z}^{(n)}$ we use 
the approach discussed in Section \ref{sec_stress_estimator} 
and the specific estimator given by equation 
\ref{equ_stress_estimator}.

For many materials, the responses of the stress component 
$\sigma_{xz}(t)$ to bulk stresses and strains 
are linear to a good approximation over a wide range of 
frequencies provided the stress and strain 
amplitudes are sufficiently small~\citep{bird1987}.  As a measure 
of the material response, we consider the dynamic complex modulus 
$G(\omega) = G'(\omega) + iG''(\omega)$, whose components are  
defined from measurements of the stress as the best least-squares fit 
of the periodic stress component $\sigma_{xz}(t)$ by the function 
$g(t) = G'(\omega) \gamma^0 \cos(\omega t) + G''(\omega) \gamma^0 \sin(\omega t)$.
This offers one characterization of the response of the material 
to oscillating bulk shear stresses and strains as the frequency
$\omega$ is varied.

To estimate the dynamic complex modulus in practice the 
least squares fit is performed for $\sigma_{xz}(t)$ over the entire
stochastic trajectory of the simulations (after some transient period).
Throughout our discussion we refer to $\theta = \omega{t}$ as the 
phase of the periodic response.  In our simulations the maximum strain 
each period was chosen to always be half the periodic unit cell in 
the x-direction, corresponding to strain amplitude $\gamma^0 = \frac{1}{2}$.  
This was achieved by adjusting the shear rate amplitude for
each frequency using $\dot{\gamma}^0 = \gamma^0\omega$.

\begin{figure}[t*]
\centering
\epsfxsize = 5in
\epsffile[-124   197   737   593]{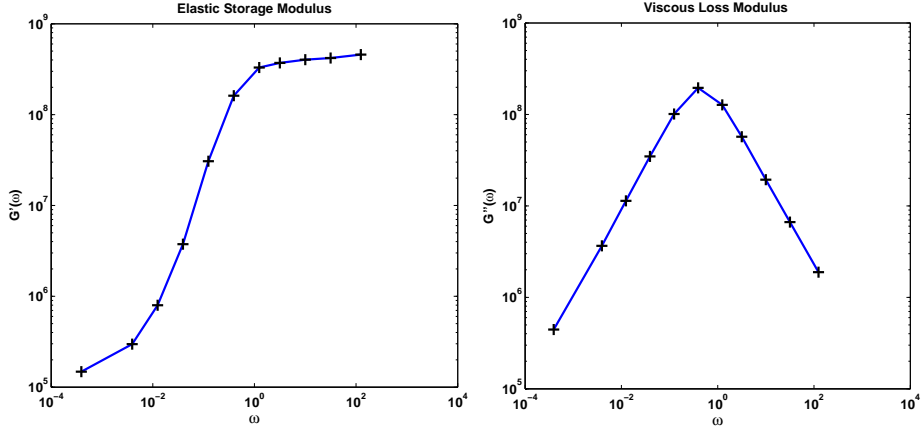}
\caption[Vesicle Response]
{Frequency response of the dynamic complex modulus of the vesicle fluid subject to 
an oscillating shear flow.  Throughout the simulations the 
total strain was held fixed to be half the domain length,
$\gamma^0 = \frac{1}{2}L$.  For a description of the parameters 
and values used in the simulations, see Table~\ref{table_vesicle_param_descr} 
and~\ref{table_vesicle_param_value}.}
\label{figure_vesicle_complex_modulus}
\end{figure}

Simulations were performed with the SELM approach in the regime 
where the hydrodynamic modes were treated as relaxed to statistical
steady-state.  The specific coupling operators $\Lambda$ and 
$\Gamma$ from~\ref{equ_op_SIB_Lambda} and~\ref{equ_op_SIB_Gamma} were used.
The simulation results of the complex modulus response of the 
vesicle when subject to a wide range of frequencies is shown 
in Figure~\ref{figure_vesicles_sheared_low_freq}, Figure~\ref{figure_vesicles_sheared_high_freq},
and Figure~\ref{figure_vesicle_complex_modulus}.  It was found that at low 
frequency the vesicle shape distortion is small and masked by thermal 
fluctuations.  At low frequency 
the vesicle membrane stresses equilibrate 
to a good approximation with the bulk shear stresses,
as illustrated in the plot of $\sigma_{xz}$ in Figure~\ref{figure_vesicles_sheared_low_freq}.  
It was found at high frequency 
the vesicle shape is visibly distorted and the membrane 
stresses do not have time to equilibrate with the bulk
shear stresses, as illustrated by the configurations 
for phase $\theta = 1.6, 0.4$ and the plot of $\sigma_{xz}$
in Figure~\ref{figure_vesicles_sheared_high_freq}.
For the vesicle configurations 
shown, the low frequency response corresponds to 
$\omega =  3.9294 \times 10^{-3} \mbox{ns}^{-1}$,
$\dot{\gamma} = 1.9647 \times 10^{-3} \mbox{ns}^{-1}$,
$\sigma_0 = 3.7114\times 10^{8} \mbox{amu}\cdot\mbox{nm}^{-1}\cdot\mbox{ns}^{-2}$,
and the 
high frequency response corresponds to 
$\omega = 1.2426 \times 10^2 \mbox{ns}^{-1}$,
$\dot{\gamma} = 6.2129\times 10^{1} \mbox{ns}^{-1}$,
$\sigma_0 = 4.6314\times 10^{10} \mbox{amu}\cdot\mbox{nm}^{-1}\cdot\mbox{ns}^{-2}$.
The phase $\theta = \omega{t}$ is reported in the range 
$[0,2\pi)$.  A description of the parameters and specific values used in 
the simulations can be found in 
Table~\ref{table_vesicle_param_descr} and~\ref{table_vesicle_param_value}.

\begin{table}[h]
\centering
\begin{tabular}{|l|l|}
\hline
Parameter & Description \\
\hline
$N$                           & Number of mesh points in each direction.\\
$\Delta{x}$                   & Mesh spacing.                  \\
$L$                           & Domain size in each direction. \\
$T$                           & Temperature.                   \\
$k_B$                         & Boltzmann's constant. \\                              
$\mu$                         & Dynamic viscosity of the solvent fluid. \\                             
$\rho$                        & Mass density of the solvent fluid. \\                             
$K_1$                         & Vesicle bond stiffness. \\
$K_2$                         & Vesicle bending stiffness. \\
$D$                           & Vesicle diameter. \\
$\omega$                      & Frequency of oscillating shearing motion. \\
$\theta$                      & Phase of the oscillatory motion, $\theta = \omega t$. \\
$\dot{\gamma}$                & Shear rate. \\
$\dot{\gamma}^0$              & Shear rate amplitude. \\
$\gamma$                      & Strain rate. \\
$\gamma^0$                    & Strain rate amplitude. \\
\hline
\end{tabular}
\caption[Vesicle Parameter Description]
{Description of the parameters used in simulations of the vesicle fluid.
\label{table_vesicle_param_descr}}
\end{table}

\begin{table}[t]
\centering
\begin{tabular}{|l|l|}
\hline
Parameter & Value \\
\hline
$N$                           & $27$ \\
$\Delta{x}$                   & $7.5 \mbox{ nm}$ \\
$L$                           & $2.025\times10^2 \mbox{ nm}$ \\
$T$                           & $300 \mbox{ K}$ \\
$k_B$                         & $8.3145 \times10^3    \mbox{ nm}^2\cdot\mbox{amu}\cdot\mbox{ns}^{-2}\cdot\mbox{K}^{-1}$ \\
$\mu$                         & $6.0221 \times 10^{5} \mbox{ amu}\cdot\mbox{cm}^{-1}\cdot\mbox{ns}^{-1}$ \\
$\rho$                        & $6.0221 \times 10^{2} \mbox{ amu}\cdot\mbox{nm}^{-3}$ \\
$\mbox{K}_1$                  & $2.2449 \times 10^{7} \mbox{ amu}\cdot \mbox{ns}^{-2}$ \\
$\mbox{K}_2$                  & $8.9796\times 10^{7}$ \\
$D$                           & $50 \mbox{ nm}$ \\
\hline
\end{tabular}
\caption[Vesicle Parameter Description]
{Fixed values of the parameters used in simulations of the vesicle fluid.
\label{table_vesicle_param_value}}
\end{table}

\clearpage
\newpage

\subsection{Application III: Rheology of a Gel-like Material}
As a further demonstration of the applicability of the SELM
approach we show how the stochastic numerical methods can be 
used to investigate properties of a gel-like material subject 
to shear.  The methods are used to study how the shear viscosity 
changes over time as the gel is subjected to shear at a constant 
rate.

\begin{figure}[t*]
\centering
\epsfxsize = 5in
\epsffile[14 14 375 137]{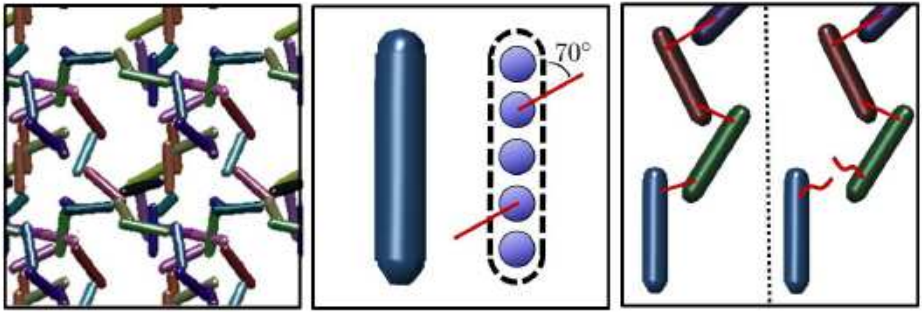}
\caption[Model for a Gel-Like Material.]
{Model for a Gel-Like Material.  The gel is formed by
polymeric chains which bind together, shown on the left.
The polymeric chains are each comprised of five control 
points and have specialized binding sites 
at the second and fourth control point, shown in the center.
The inter-polymer bonds have a preferred extension and angle.
When an inter-polymer bond is strained beyond $50\%$ of its
preferred rest length the bond breaks irreversibly, shown
on the right.
}
\label{figure_gelSchematic}
\end{figure}

The gel-like material is modeled as a collection of 
polymer chains which are able to bond together 
at two specialized sites along the chain, see Figure~\ref{figure_gelSchematic}. 
The energy associated with the mechanics of the individual
polymer chains and the bonds which they form are given by
\begin{eqnarray}
\phi_{1}(r)       & = & \frac{1}{2}K_{1} (r - r_{0,1})^2 \\
\phi_{2}(\boldsymbol{\tau}_1,\boldsymbol{\tau}_2) 
                  & = &  
\frac{1}{2} K_{2}
\left|\boldsymbol{\tau}_{1} - \boldsymbol{\tau}_{2}\right|^2 \\
\phi_{3}(r)       & = & \sigma^2 K_{3}\exp\left[-\frac{(r - r_{0,3})^2}{2\sigma^2}\right] \\
\phi_{4}(\theta)  & = & -K_{4} \cos(\theta - \theta_{0,4}).
\end{eqnarray}
The $r$ is the separation distance between two control points,
$\theta$ is the bond angle between three control points, 
and $\boldsymbol{\tau}$ is a tangent vector along the polymer 
chain, see Figure ~\ref{figure_gelSchematic}.

The $\phi_1$ energy accounts for stretching of a bond within a polymer chain from its preferred
extension $r_{0,1}$. The $\phi_2$ energy accounts for bending of the polymer chain locally.
To account for interactions at the specialized binding sites of the polymers the potentials
$\phi_3$ and $\phi_4$ are introduced.  The potential $\phi_3$ gives the energy of the 
bond between the two polymer chains and penalizes deviation from the preferred bond extension
$r_{0,3}$.  The exponential of $\phi_3$ is introduced so that the resistance in the 
bond behaves initially like a harmonic bond but decays rapidly to zero when the bond is 
stretched beyond the length $\sigma$.  The potential $\phi_4$ 
gives the energy for the preferred bond angle when two of the polymer 
chains are bound together.

The total energy of the system is given by 
\begin{eqnarray}
\Phi[\mathbf{X}] & = & E_1[\mathbf{X}] + E_2[\mathbf{X}] + E_3[\mathbf{X}] + E_4[\mathbf{X}] \\
E_1[\mathbf{X}]  & = & \sum_{(i,j)   \in \mathcal{Q}_1}  \phi_1(r_{ij}), \mbox{\hspace{0.25cm}}
E_2[\mathbf{X}]   =  \sum_{(i,j,k) \in \mathcal{Q}_2} \phi_2(\boldsymbol{\tau}_{ij},\boldsymbol{\tau}_{jk}) \\
E_3[\mathbf{X}]  & = & \sum_{(i,j)   \in \mathcal{Q}_3}  \phi_3(r_{ij}), \mbox{\hspace{0.25cm}}
E_4[\mathbf{X}]   =  \sum_{(i,j,k) \in \mathcal{Q}_4} \phi_4(\theta_{ijk}).
\end{eqnarray}
The sets $\mathcal{Q}_k$ define the interactions according to the 
structure of the individual polymer chains and the topology of 
the gel network.  When bonds are stretched beyond the critical length $3\sigma$ they 
are broken irreversibly, which results in the sets $\mathcal{Q}_3$ and $\mathcal{Q}_4$
being time dependent.

\begin{figure}[t*]
\centering
\epsfxsize = 3in
%\epsffile[86 228 527 563]{compareShearViscosityFENE3.eps}
\epsffile[81   227   529   564]{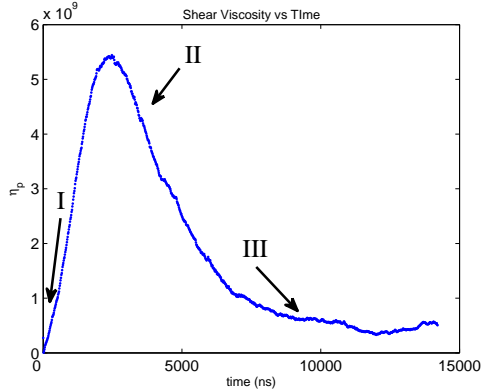}
\caption[Shear Viscosity of a Gel-like Material]
{Study of the shear viscosity of a gel-like material.  At time
zero the material has weak bonds between short polymeric
chains.  Under the shear deformation the gel is stretched 
and the bonds are strained until ultimately breaking.  Many of
the polymers are misaligned with the direction of fluid
flow and are further stretched by the fluid shear stresses.
As the polymer chains align with the direction of fluid flow
the forces transmitted in the direction of shear decrease
and the shear viscosity approaches a steady-state value.
The thermal fluctuations maintain transient misalignments
of the polymer chains which transmit forces in the direction 
of shear resulting in a contribution to the shear viscosity
which is non-zero at steady-state.  The microstructure 
reordering in each of these stages, labeled I, II, III, 
is reflected in the shear viscosity of the material as 
a function of time and in Figure~\ref{figure_thixotropyGelLikeMaterial_configComposite}.
For the specific physical parameters used in these simulation 
see Table \ref{table_gel_param_descr} and \ref{table_gel_param_value}.
}
\label{figure_thixotropyGelLikeMaterial}
\end{figure}

\begin{figure}[t*]
\centering
\epsfxsize = 5in
%\epsffile[-246   218   858   572]{fig_thixotropyGelMaterial_configComposite.eps}
\epsffile[14 14 375 141]{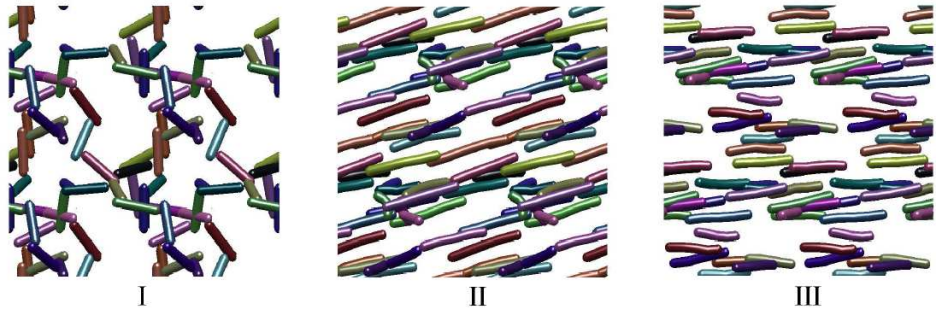}
\caption[Microstructure of the Gel-like Material]
{The microstructure of a gel-like material at three 
different times.  On the left is shown the microstructure
of the gel-like material before any shear has been applied.
In the middle is show the microstructure of the gel after
almost all of the bonds between polymer chains have been
broken.  In this case, the misaligned polymer chains 
continue to be stretched by the shear stresses of the 
fluid yielding a relatively large effective shear viscosity.  On the 
right is shown the microstructure of the gel when the 
system has relaxed to statistical steady-state.  In this case,
the thermal fluctuations drive transient misalignments
of the polymer chains with the direction of flow 
which on average make a non-zero contribution to
the shear viscosity.  The times shown in each of these 
figures is $t = 0\mbox{ ns}$, $t = 2844\mbox{ ns}$, $t = 7111\mbox{ ns}$.
For the specific physical parameters used in these simulations 
see Table \ref{table_gel_param_descr} and \ref{table_gel_param_value}.
}
\label{figure_thixotropyGelLikeMaterial_configComposite}
\end{figure}

\begin{table}[t]
\centering
\begin{tabular}{|l|l|}
\hline
Parameter & Description \\
\hline
$N$                             & Number of mesh points in each direction.\\
$\Delta{x}$                     & Mesh spacing.                  \\
$\Delta{t}$                     & Time step.                     \\
$L$                             & Domain size in each direction. \\
$T$                             & Temperature.                   \\
$k_B$                           & Boltzmann's constant. \\                              
$\mu$                           & Dynamic viscosity of the solvent fluid. \\                             
$\rho$                          & Mass density of the solvent fluid. \\                             
$\dot{\gamma}$                  & Shear rate. \\
$N_p$                           & Number of polymer chains. \\
$N_s$                           & Number of control points per polymer chain. \\
$r_p$                           & Polymer effective cylindrical radius. \\
$K_{1}$                         & Stiffness of the bonds of the polymer chain. \\
$r_{0,1}$                       & Rest length of the bonds of the polymer chain. \\
$K_{2}$                         & Bending stiffness of the polymer chain. \\
$K_{3}$                         & Stiffness of the bonds at a polymer binding site. \\
$r_{0,3}$                       & Rest length of the bond at a polymer binding site. \\
$K_{4}$                         & Bending stiffness of the bond at a polymer binding site. \\
$\theta_{0,4}$                  & Preferred angle of a bond at a polymer binding site. \\
\hline
\end{tabular}
\caption[Gel-Like Material Parameter Description]
{Description of the parameters used in simulations of the gel-like material.
\label{table_gel_param_descr}}
\end{table}

\begin{table}[t]
\centering
\begin{tabular}{|l|l|}
\hline
Parameter & Value \\
\hline
$N$                           & $72$ \\
$\Delta{x}$                   & $11.25  \mbox{ nm}$ \\
$\Delta{t}$                   & $1.4222 \mbox{ ns}$ \\
$L$                           & $810 \mbox{ nm}$ \\
$T$                           & $300 \mbox{ K}$ \\
$k_B$                         & $8.3145 \times10^3    \mbox{ nm}^2\cdot\mbox{amu}\cdot\mbox{ns}^{-2}\cdot\mbox{K}^{-1}$ \\
$\mu$                         & $6.0221 \times 10^{5} \mbox{ amu}\cdot\mbox{cm}^{-1}\cdot\mbox{ns}^{-1}$ \\
$\rho$                        & $6.0221 \times 10^{2} \mbox{ amu}\cdot\mbox{nm}^{-3}$ \\
$\dot{\gamma}$                & $1.2 \times 10^{-3} \mbox{ ns}^{-1}$ \\
$N_p$                         & $110$ \\
$N_s$                         & $5$ \\
$r_p$                         & $15 \mbox{ nm}$ \\
$K_{1}$                       & $2.9932 \times 10^{5} \mbox{ amu}\cdot\mbox{ns}^{-2}$  \\
$r_{0,1}$                     & $30 \mbox{ nm}$ \\
$K_{2}$                       & $2.9932 \times 10^{8}$ \\
$K_{3}$                       & $2.9932 \times 10^{5} \mbox{ amu}\cdot\mbox{ns}^{-2}$  \\
$r_{0,3}$                     & $30 \mbox{ nm}$ \\
$K_{4}$                       & $2.9932 \times 10^{8}$ \\
$\theta_{0,4}$                & $70\,^{\circ}$ \\
\hline
\end{tabular}
\caption[Gel-like Material Parameter Description]
{Fixed values of the parameters used in simulations of the gel-like material.
\label{table_gel_param_value}}
\end{table}

To study the rheological response of the gel-like material 
the system is subjected to shear at a constant rate.  
To obtain an effective macroscopic stress $\sigma_p$ 
for the system the estimator is used from equation~\ref{equ_stress_estimator}.
To characterize the rheological response 
we use the shear viscosity defined by 
\begin{eqnarray}
\eta_p & = & {\sigma_p^{(s,v)}}/{\dot{\gamma}}.
\end{eqnarray}
The $\dot{\gamma}$ is the rate of shear of the polymeric fluid.
In the notation, the superscript $(s,v)$ indicates the 
tensor component with the index $s$ corresponding to the 
direction of shear and the index $v$ corresponding to 
the direction of the fluid velocity.  The contributions of the 
solvent fluid to the shear viscosity can be considered 
separately~\citep{Bird1987Vol2}.

The entire gel network experiences an unbounded shear 
deformation.  This is expected to result in breakage of 
bonds of the gel network.  This suggests that the 
rheological response will depend on how long the 
material has been subject to shear.  To investigate 
the role reorganization at the microstructure level,
repeated stochastic simulations are carried out 
using the SELM approach to determine the 
effective shear viscosity of the material
as a function of time.

An interesting behavior is found in which the 
material initially exhibits an increased shear 
viscosity before settling down to a steady-state
value.  The responses of the material to shear
can be roughly divided into three stages.
In the first, there is an initial increase which
can be attributed to the stretching of the 
inter-chain bonds between the polymer chains 
and the intra-chain bonds within each polymer 
chain, which occurs as the gel as a whole is 
strained.  After a relatively short period, 
the bonds between the polymer chains are
observed to break with the remaining stress arising
from the stretching of the polymer chains which 
occurs from the shear stresses of the fluid and 
misalignment with the direction of flow, see 
the region labeled by I in Figure~\ref{figure_thixotropyGelLikeMaterial} 
and~\ref{figure_thixotropyGelLikeMaterial_configComposite}.

In the second stage, the individual polymer chains 
rotate and begin to align with the direction of flow.
As a result of the intra-chain restoring
forces the strain of the individual polymer chains
is reduced.  The increased alignment and reduced
strain of the polymer chains yields an overall 
decrease in the forces transmitted in the direction 
of shear.  Consequently, the shear viscosity 
begins to decrease, see 
the region labeled by II in Figure~\ref{figure_thixotropyGelLikeMaterial} 
and~\ref{figure_thixotropyGelLikeMaterial_configComposite}.

In the last stage, the chains eventually settle 
into a statistical steady-state in which the 
thermal fluctuations drive the chains to 
misalign only transiently with the flow direction.  
These misaligned excursions by the polymer chains 
sustained by the thermal fluctuations result 
in forces transmitted in the direction of 
shear on average.  This is reflected in the shear viscosity 
by a non-zero steady-state value, see 
the region labeled by III in Figure~\ref{figure_thixotropyGelLikeMaterial} 
and~\ref{figure_thixotropyGelLikeMaterial_configComposite}.

Using the SELM approach more complicated 
situations could also be studied,
such as the case in which the 
bonds between the polymer chains are able 
to reform.  An interesting investigation 
in this case would be to study how the 
viscosity behaves after decreasing or 
ceasing shearing of the system for a 
period of time.  In this case the gel 
would have time to reform structures 
before being again subjected to large
shears.  Using such a SELM approach a 
widely variety of shear thinning and 
thixotropic phenomena could be studied 
at the level of the 
microstructures~\citep{Barnes1997,Bird1987Vol2,Doi1986}.

\clearpage
\newpage

\section{Conclusions}

A general formalism was developed which allows for
the coupling of Eulerian and Lagrangian 
descriptions of physical systems.   A general approach 
was introduced for incorporating thermal fluctuations 
in such descriptions.  The approach addresses both 
the inertial regime and the overdamped regime.  For 
the study of rheological responses of 
materials, an approach was developed which allows 
for generalized periodic boundary conditions which 
induce the shear.  For simulations using the formalism 
stochastic numerical methods were developed which 
efficiently generate the required stochastic driving
fields.  As a demonstration of how these methods can be
used in practice, simulation studies were carried out 
for complex fluids and soft materials.
The basic Stochastic Eulerian Lagrangian Method (SELM) 
approach is expected to be useful in the formulation 
of descriptions and computational approaches
for the study of a wide variety of fluid 
structure phenomena involving thermal fluctuations.
\\

\section{Acknowledgements}
The author P.J.A. acknowledges support from research grant 
NSF DMS-0635535.  We would especially like to
thank Aleksandar Donev, Alejandro Garcia,
John Bell, and Tony Ladd for stimulating 
conversations about this work. 
This paper is dedicated in memorial to Tom Bringley,
whose academic publications continue to inspire.
His passion for life, mathematics, and science
will be greatly missed.

\bibliographystyle{siam}
\bibliography{finalManuscript_submit}

\appendix

\section{Invariance of the Boltzmann Distribution under SELM Stochastic Dynamics}
\label{appendix_invariance_Boltzmann}

The probability distribution of the stochastic 
equations~\ref{equ_EL_E}-\ref{equ_EL_L}
are governed formally by the
Fokker-Planck equation
\begin{eqnarray}
\label{equ_Fokker_Planck_abs}
\frac{\partial \Psi}{\partial t} = -\nabla \cdot \mathbf{J}
\end{eqnarray}
with the probability flux given by
\begin{eqnarray}
\mathbf{J} & = &
\left[
\begin{array}{l}
\mathcal{L}\mathbf{p} \Psi + (\Lambda + \lambda)\Psi -\frac{1}{2} G \nabla_{\mathbf{p}} \Psi \\
(\Gamma + \gamma) \Psi -\frac{1}{2} W \nabla_{\mathbf{X}}\Psi
\end{array}
\right].
\end{eqnarray}
The $G$, $W$ are the covariance operators associated with $\mathbf{g}$ and $\mathbf{Z}$.
The $\Psi(\mathbf{p},\mathbf{X},t)$ is the formal probability density for finding the 
system in state $(\mathbf{p},\mathbf{X})$ at time $t$.
For the present purposes our discussion will only be formal since the 
SPDEs are infinite dimensional and for the density there is no Lebesgue measure for  
the function space, see~\citep{Oksendal2000, Gardiner1985, Da1992}.  In practice
a finite dimensional stochastic process will always be used to approximate the 
SPDEs and has a probability distribution satisfying a well-defined equation.

For the systems under consideration, the Boltzmann distribution has the form
$\Psi_{BD}(\mathbf{p},\mathbf{X}) = \frac{1}{Z} \exp\left[-{E[\mathbf{p},\mathbf{X}]}/{k_B T}\right]$,
where $Z$ is a normalization constant so that $\Psi_{BD}$ integrates to one~\citep{Reichl1998}.  
The requirement that this distribution is invariant 
under the stochastic dynamics of~\ref{equ_EL_E}-\ref{equ_EL_L} is equivalent 
to $\nabla \cdot \mathbf{J} = 0$.  This requires
\begin{eqnarray}
\label{equ_appendix_div_J}
\nabla \cdot \mathbf{J} & = & A_1 + A_2 + \nabla\cdot \mathbf{A}_3 = 0 \\
\label{equ_appendix_A_1}
A_1            & = & \left[(\Lambda + \lambda)\cdot\nabla_{\mathbf{p}}E + (\Gamma + \gamma)\cdot\nabla_{\mathbf{X}}E
\right](-k_B{T})^{-1}\Psi \\
\label{equ_appendix_A_2}
A_2            & = & \left(\nabla_{\mathbf{p}}\cdot(\Lambda + \lambda) 
+ \nabla_{\mathbf{X}}\cdot(\Gamma + \gamma)\right)\Psi \\
\label{equ_appendix_A_3}
\mathbf{A}_3   & = & \left(\mathcal{L}\mathbf{p} + \frac{G\nabla_{\mathbf{p}}E + W\nabla_{\mathbf{X}}E}{2k_B{T}}\right)\Psi.
\end{eqnarray}
For the energy given by equation~\ref{equ_SELM_energy} we have 
\begin{eqnarray}
\label{equ_appendix_grad_p_E}
\nabla_{\mathbf{p}}E & = & \rho_0^{-1}\mathbf{p} \\
\label{equ_appendix_grad_X_E}
\nabla_{\mathbf{X}}E & = & \nabla_{\mathbf{X}}\Phi = -\mathbf{F} 
\end{eqnarray}
where $\mathbf{F}$ denotes the force for the configuration.

Now we can derive conditions for the coupling operators by requiring
that $A_1 = A_2 = 0$ for all possible values of 
$\mathbf{p}$ and $\mathbf{F}$.  The requirement that  
$A_1 = 0$ corresponds to the energy being
conserved under the dynamics of equations 
\ref{equ_EL_E}-\ref{equ_EL_L} when 
$\mathbf{g} = \mathbf{Z} = 0$ and 
$\boldsymbol{\sigma} = 0$.  For 
these dynamics the energy satisfies
${dE}/{dt} = (\Lambda + \lambda)\cdot\nabla_{\mathbf{p}}E + (\Gamma + \gamma)\cdot\nabla_{\mathbf{X}}E 
= 0$.  Since the forces associated with time independent constraints do not do any work 
on the system we have
that $\lambda\cdot\nabla_{\mathbf{p}}E + \gamma\cdot\nabla_{\mathbf{X}}E 
= 0$.  Conservation of energy then requires  
$\Lambda\cdot\nabla_{\mathbf{p}}E + \Gamma\cdot\nabla_{\mathbf{X}}E = 0$. 
By using the variational derivatives~\citep{Gelfand2000} of $E$ given in 
\ref{equ_appendix_grad_p_E}-\ref{equ_appendix_grad_X_E} we have
$\Lambda\cdot\nabla_{\mathbf{p}}E = \int \Lambda \rho_0^{-1}\mathbf{p} d\mathbf{x}$
and $\Gamma\cdot\nabla_{\mathbf{X}}E = \int -\Gamma \mathbf{F} d\mathbf{q}$.  
By substituting these expressions 
into \ref{equ_appendix_A_1}, we obtain from $A_1 = 0$ that 
the condition~\ref{equ_coupling_cond_energy} must be satisfied.

The requirement that $A_2 = 0$ requires that 
the dynamical flow in phase space defined by 
$(\Lambda + \lambda,\Gamma + \gamma)$ 
is volume preserving.  For the dynamics when 
$\mathbf{g} = \mathbf{Z} = 0$, $\boldsymbol{\sigma} = 0$,
and $E = 0$ this condition is equivalent to requiring that 
the uniform distribution is invariant under the dynamics.  
The condition~\ref{equ_coupling_cond_uniform} follows
by using the function representing the variational derivatives~\citep{Gelfand2000} 
appearing in the divergence operation corresponds to
$\nabla_{\mathbf{X}}\cdot\Gamma 
= \int (\delta \Gamma / \delta \mathbf{X})(\mathbf{q},\mathbf{q}) d\mathbf{q}$,
$\nabla_{\mathbf{X}}\cdot\gamma 
= \int (\delta \gamma / \delta \mathbf{X})(\mathbf{q},\mathbf{q}) d\mathbf{q}$,
and similarly for $\Lambda$, $\lambda$.

The requirement that $A_3 = 0$ requires from 
equation \ref{equ_appendix_grad_p_E}-\ref{equ_appendix_grad_X_E}
that \\
$\mathcal{L}\mathbf{p} + \left[{(G\rho^{-1}\mathbf{p} - W\mathbf{F})}/{2k_B{T}}\right] = 0$
for any $\mathbf{p}$ and $\mathbf{F}$.  This requirement corresponds to
the condition of Detailed-Balance of statistical mechanics~\citep{Reichl1998}.
Since $\mathbf{p}$ and $\mathbf{F}$ are arbitrary,
this requires that $W = 0$ so that $\mathbf{Z} = 0$.  This also requires that 
$G = -2\mathcal{L}\mathcal{C}$ with $\mathcal{C} = k_B{T}\rho_0\mathcal{I}$,
where $\mathcal{I}$ is the identity operator.  This yields condition~\ref{equ_SELM_covOp_G}.
From the form of the energy in~\ref{equ_SELM_energy} and the Boltzmann distribution
we see the equilibrium fluctuations of $\mathbf{p}$ are Gaussian
with covariance $\mathcal{C}$.  This condition relates the equilibrium
fluctuations to the dissipative operator of the system and  
is a variant of the Fluctuation-Dissipation Principle of statistical 
mechanics~\citep{Reichl1998}.  This shows that provided the 
coupling operators and stochastic fields satisfy conditions 
\ref{equ_coupling_cond_energy}, \ref{equ_coupling_cond_uniform}, and
\ref{equ_SELM_covOp_G}, the Boltzmann distribution is invariant 
under the SELM stochastic dynamics.

For the discretized equations, we now derive conditions 
\ref{equ_coupling_cond_energy_discr}, \ref{equ_coupling_cond_uniform_discr}, and
\ref{equ_SELM_covOp_G_discr}.  The calculations 
follow similarly to the case above so we only state the basic features
of the derivation.  For the discretized equations
the probability flux is given by
\begin{eqnarray}
\mathbf{J} & = &
\left[
\begin{array}{l}
L\mathbf{p} \Psi + (\Lambda + \lambda)\Psi -\frac{1}{2} G \nabla_{\mathbf{p}} \Psi \\
(\Gamma + \gamma) \Psi
\end{array}
\right]
\end{eqnarray}
where $\mathbf{p}$ and $\mathbf{X}$ are now finite dimensional vectors.
The Boltzmann distribution now uses the energy of the discrete system
\begin{eqnarray}
E[\mathbf{p},\mathbf{X}] 
= \sum_{\mathbf{m}} \frac{1}{2} \rho_{0}^{-1} |\mathbf{p}_{\mathbf{m}}|^2 \Delta{x}^d 
+ \Phi(\mathbf{X})
\end{eqnarray}
with
\begin{eqnarray}
\label{equ_appendix_grad_p_E_discr}
\nabla_{\mathbf{p}}E & = & \rho_0^{-1}\mathbf{p} \Delta{x}^d \\
\label{equ_appendix_grad_X_E_discr}
\nabla_{\mathbf{X}}E & = & \nabla_{\mathbf{X}}\Phi = -\mathbf{F}. 
\end{eqnarray}
Substituting these expressions in 
\ref{equ_appendix_div_J} - \ref{equ_appendix_A_3} and reasoning as above
yields the conditions \ref{equ_coupling_cond_energy_discr}, 
\ref{equ_coupling_cond_uniform_discr}, and \ref{equ_SELM_covOp_G_discr}.

\section{A Fluctuation-Dissipation Principle for Time-Dependent Operators}
\label{appendix_fluct_dissip}
Consider the stochastic process given by
\begin{eqnarray}
d\mathbf{z}_t & = & L(t)\mathbf{z} dt + Q(t)d\mathbf{B}_t \\
G(t)          & = & QQ^T.
\end{eqnarray}
We now establish the following fluctuation-dissipation relation
\begin{eqnarray}
\label{equ_G_t_fluct_dissp}
G(t) = -L(t)\bar{C} - \bar{C}^TL(t)^T.
\end{eqnarray}
This relates the covariance $G(t)$ of the stochastic 
driving field to a time-dependent dissipative operator 
$L(t)$ and a time-independent equilibrium covariance 
$\bar{C}$.  We show that this relation allows for 
$G(t)$ to be chosen to ensure that the stochastic 
dynamics exhibits at statistical steady-state 
equilibrium fluctuations with the specified
covariance $\bar{C}$.

Let the covariance at time $t$ be denoted by
\begin{eqnarray}
C(t) = \langle\mathbf{u}(t)\mathbf{u}(t)^T\rangle.
\end{eqnarray}
By Ito's Lemma the second moment satisfies
\begin{eqnarray}
\label{equ_M_t}
dC(t) = \left(L(t)C(t) + C(t)^TL(t)^T + G(t)\right) dt.
\end{eqnarray}
It will be convenient to express this equation 
by considering all of the individual entries of the 
matrix $C(t)$ collected into a single column vector 
denoted by $\mathbf{c}_t$.  Similarly, for covariance 
matrix $G(t)$ we denote the column vector of entries
by $\mathbf{g}_t$ and for $\bar{C}$ by 
$\bar{\mathbf{c}}$.  Since the products $L(t)C(t)$
and $C(t)^TL(t)^T$ are both linear operations in 
the entries of the matrix $C(t)$ we can express 
this in terms of multiplication by of a matrix 
$A(t)$ acting on $\mathbf{c}_t$.  

This notation allows for equation~\ref{equ_M_t} to be 
expressed equivalently as
\begin{eqnarray}
\label{equ_m_t}
d\mathbf{c}_t = \left(A(t) \mathbf{c}_t + \mathbf{g}_t\right) dt.
\end{eqnarray}
The equation~\ref{equ_M_t} can be solved formally  
by the method of integrating factors to obtain
\begin{eqnarray}
\label{equ_m_t_int}
\mathbf{c}_t = e^{\Xi(0,t)}\mathbf{c}_0
+ \int_0^t e^{\Xi(s,t)} \mathbf{g}_s ds
\end{eqnarray}
where $\Xi(s,t) = \int_s^t A(r) dr$.

The fluctuation-dissipation relation given by 
equation~\ref{equ_G_t_fluct_dissp} is equivalent to choosing 
\begin{eqnarray}
\mathbf{g}_s = -A(s)\bar{\mathbf{c}}.
\end{eqnarray}
For this choice, a useful identity is 
\begin{eqnarray}
e^{\Xi(s,t)} \mathbf{g}_s = \frac{\partial}{\partial s} e^{\Xi(s,t)}\bar{\mathbf{c}}.
\end{eqnarray}
Substitution into equation~\ref{equ_m_t_int} gives
\begin{eqnarray}
\label{equ_m_t_int2}
\mathbf{c}_t = e^{\Xi(0,t)}\mathbf{c}_0
+ \left(e^{\Xi(t,t)} - e^{\Xi(0,t)}\right) \bar{\mathbf{c}}.
\end{eqnarray}

Now, if $L(t)$ is negative definite uniformly in time, 
$\mathbf{v}^T L(t) \mathbf{v} < \alpha_0 < 0$,
then $A(t)$ is also uniformly negative definite.
This implies that 
\begin{eqnarray}
\label{equ_lim_e_Xi}
\lim_{t\rightarrow \infty} e^{\Xi(0,t)} = 0.
\end{eqnarray}
Taking the limit of both sides of equation~\ref{equ_m_t_int2}
and using equation~\ref{equ_lim_e_Xi} yields
\begin{eqnarray}
\label{equ_m_t_limit}
\lim_{t\rightarrow \infty} \mathbf{c}_t 
= \mathbf{\bar{c}}.
\end{eqnarray}

This shows that the stochastic driving field with 
covariance given by equation~\ref{equ_G_t_fluct_dissp} 
yields equilibrium fluctuations with covariance $\bar{C}$.
This extends the fluctuation-dissipation relation
to the case of time-dependent operators.

For the discretization given in Section~\ref{sec_soft_materials_subject_shear}, 
we point out some of the properties of the 
specific matrix $L(t)$ which are used.  From 
equation~\ref{equ_dw_dt_time_op} the non-zero eigenvalues of 
$L(t)$ can be shown to be negative and uniformly 
bounded away from zero in time.  The eigenvector 
associated with the zero eigenvalue of $L(t)$ is 
in fact the same for all times.  The eigenvector
of the zero eigenvalue is proportional to the 
vector with all components set to one.  In 
practice, this mode is set to zero. By 
conservation of momentum of the fluid body 
as a whole, this mode remains zero when subject to internal
conservative forces.  This allows for the operator $L(t)$
to be considered as acting on the linear space 
which excludes this null eigenvector.  On this linear space,
$L(t)$ is strictly negative definite uniformly in time.  
Similar considerations can be made when considering
the effect of the incompressibility constraint for
the operator $\tilde{L}(t) = \wp L(t)$.  Thus the 
time-dependent fluctuation-dissipation relation 
given by equation~\ref{equ_G_t_fluct_dissp} still 
holds provided the stochastic process is considered 
on the appropriate linear space which excludes the null 
eigenvectors.

\section{The Particle Representation Function $\delta_a$}
\label{appendix_delta_func}
In the immersed boundary method, it is required
that a function $\delta_a$ be specified to represent the 
elementary 
particles.  The representation of this 
function is 
often derived from the following function $\phi$ 
which is known to have desirable numerical properties 
~\citep{Peskin2002, Atzberger2007a}: 
\begin{eqnarray}
\label{equ_phi_def}
\phi(r) & = & \left\{
\begin{array}{ll}
0
                  & \mbox{, if $r \leq -2$} \\
                  & \\
\frac{1}{8} \left(5 + 2r - \sqrt{-7 - 12r - 4r^2} \right) 
                  & \mbox{, if $-2 \leq r \leq -1$} \\
                  & \\
\frac{1}{8} \left(3 + 2r + \sqrt{1 - 4r - 4r^2} \right) 
                  & \mbox{, if $-1 \leq r \leq 0$} \\
                  & \\
\frac{1}{8} \left(3 - 2r + \sqrt{1 + 4r - 4r^2} \right) 
                  & \mbox{, if $0 \leq r \leq 1$} \\
                  & \\
\frac{1}{8} \left(5 - 2r - \sqrt{-7 + 12r - 4r^2} \right) 
                  & \mbox{, if $1 \leq r \leq 2$} \\
                  & \\
0
                  & \mbox{, if $2 \leq r$.} 
                  \\
\end{array}
\right.
\end{eqnarray}

For three dimensional systems the function 
$\delta_a$ representing elementary 
 particles of size $a$ is  
\begin{eqnarray}
\label{equ_phi_def_delta_a}
\delta_a(\mathbf{r}) = \frac{1}{a^3}
\phi\left(\frac{\mathbf{r}^{(1)}}{a}\right)
\phi\left(\frac{\mathbf{r}^{(2)}}{a}\right)
\phi\left(\frac{\mathbf{r}^{(3)}}{a}\right), 
\end{eqnarray}
where the superscript indicates the index of the vector component.

To maintain good numerical properties, 
 the particles are 
restricted to sizes $a = n\Delta{x}$, where $n$ is a positive 
integer.  For a derivation and a detailed discussion of 
the properties of these functions 
see~\citep{Peskin2002, Atzberger2007a}.

\section{Table}
\vspace{0.5cm}
\label{appendix_rough_calc}

\begin{tabular}{|l|l|}
\hline
Parameter & Description \\
\hline
$N_A$                         & Avogadro's number. \\
\mbox{amu}                    & Atomic mass unit. \\
\mbox{nm}                     & Nanometer. \\
\mbox{ns}                     & Nanosecond. \\
$k_B$                         & Boltzmann's Constant. \\
$T$                           & Temperature. \\
$\eta$                        & Dynamic viscosity of water. \\
$\gamma_s = 6\pi \eta R$        & Stokes' drag of a spherical particle. \\
\hline
\end{tabular}
\vspace{1cm}

\begin{tabular}{|l|l|}
\hline
Parameter & Value \\
\hline
$N_A$                         & $6.02214199\times 10^{23}$. \\
\mbox{amu}                    & $1/10^3 N_A$ \mbox{kg}. \\
\mbox{nm}                     & $10^{-9}\hspace{0.1cm} \mbox{m}$. \\
\mbox{ns}                     & $10^{-9}\hspace{0.1cm} \mbox{s}$. \\
$k_B$                         & $8.31447\times 10^{3} \hspace{0.1cm} \mbox{amu}\hspace{0.1cm}\mbox{nm}^2/\mbox{ns}^2\hspace{0.1cm}K$. \\
$T$                           & $300\mbox{K}$. \\
$\eta$                        & $6.02214199\hspace{0.1cm} \mbox{amu}/\mbox{cm}\hspace{0.1cm}\mbox{ns}$. \\
\hline
\end{tabular}
\vspace{1cm}

\end{document}